\documentclass[sn-mathphys-num]{sn-jnl}

\usepackage{graphicx}%
\usepackage{multirow}%
\usepackage{amsmath,amssymb,amsfonts}%
\usepackage{amsthm}%
\usepackage{mathrsfs}%
\usepackage[title]{appendix}%
\usepackage{xcolor}%
\usepackage{textcomp}%
\usepackage{manyfoot}%
\usepackage{booktabs}%
\usepackage{algorithm}%
\usepackage{algorithmicx}%
\usepackage{algpseudocode}%
\usepackage{listings}%
\usepackage{subfig}%
\usepackage{subcaption}%
\usepackage{sidecap}%
\usepackage{geometry}%
\usepackage[numbers]{natbib}%
\usepackage{cleveref}%

\crefname{figure}{Fig.}{Fig.} 

\begin{document}
	\newgeometry{margin=2cm, height=10in, includefoot}
	\title[Article Title]{Investigation of Low Frequency Noise in CryoCMOS devices through Statistical Single Defect Spectroscopy}
	
	
	\author*[1,2]{\fnm{E.} \sur{Catapano}}\email{edoardo.catapano@imec.be}
	
	\author[1,3]{\fnm{A.} \sur{Varanasi}}
	\author[1]{\fnm{P.} \sur{Roussel}}
	\author[1]{\fnm{R.} \sur{Degraeve}}
	\author[1]{\fnm{Y.} \sur{Higashi}}
	\author[1]{\fnm{R.} \sur{Asanovski}}
	\author[1]{\fnm{B.} \sur{Kaczer}}
	\author[1]{\fnm{J.} \sur{Diaz Fortuny}}
	\author[4]{\fnm{M.} \sur{Waltl}}
	\author[1,2]{\fnm{V.} \sur{Afanasiev}}
	\author[1,5]{\fnm{K.} \sur{De Greve}}
	\author[1]{\fnm{A.} \sur{Grill}}
	
	\affil*[1]{\orgname{imec}, \city{Heverlee}, \postcode{3001}, \country{Belgium}}
	
	\affil[2]{\orgdiv{Department of Physics and Astronomy}, \orgname{KU Leuven}, \city{Heverlee}, \postcode{3001}, \country{Belgium}}
	
	\affil[3]{\orgdiv{Department of Materials Engineering}, \orgname{KU Leuven}, \postcode{3001} Leuven, \country{Belgium}}
	
	\affil[4]{\orgdiv{Christian Doppler Laboratory for Single-Defect Spectroscopy in Semiconductor Devices at the Institute for Microelectronics}, \orgname{TU Wien}, \city{Vienna}, \country{Austria}}
	
	\affil[5]{\orgdiv{Department of Electrical Engineering, ESAT-MNS and Proximus Chair in Quantum Science and Technology}, \orgname{KU Leuven}, \city{Heverlee}, \postcode{3001}, \country{Belgium}}

	\abstract{High 1/f noise in CryoCMOS devices is a critical parameter to keep under control in the design of complex circuits for low temperatures applications. Current models predict the 1/f noise to scale linearly with temperature, and gate oxide defects are expected to freeze out at cryogenic temperatures. Nevertheless, it has been repeatedly observed that 1/f noise deviates from the predicted behaviour and that gate oxide defects are still active around 4.2 K, producing random telegraph noise. In this paper, we probe single gate oxide defects in 2500 nMOS devices down to 5 K in order to investigate the origin of 1/f noise in CryoCMOS devices. From our results, it is clear that the number of defects active at cryogenic temperatures resulting in random telegraph noise is larger than at 300 K. Threshold voltage shifts due to charged defects are shown to be exponentially distributed, with different modalities across temperatures and biases: from monomodal at 300 K to trimodal below 100 K. Distributions' shapes are interpreted in the framework of percolation theory. By fitting these distributions, it is shown that more than 80\% of the detected defects belongs to the oxide bulk. Afterwards, starting from the raw data in time domain, we reconstruct the low frequency noise spectra, highlighting the contributions of defects belonging to different branches and, therefore, to different oxide layers. This analysis shows that, although interface traps and large defects associated with the third mode are the main sources of 1/f noise at 5 K, bulk oxide defects still contribute significantly to low-frequency noise at cryogenic temperatures. Finally, we show that defect time constants and step heights are uncorrelated, proving that elastic tunnelling model for charge trapping is not accurate.}


	\maketitle
	
	\section{Introduction}\label{sec1}
	
	 In recent years, technologies related to quantum computing progressed considerably, in terms of both qubits quality and performance \cite{van_damme_advanced_2024} and of co-integration with classical electronics for qubits control and read-out \cite{xue_cmos-based_2021, acharya_multiplexed_2023}. Since in most of the present technologies the control read-out electronics is placed at room temperature, outside the cryogenically cooled environment (dilution refrigerator) of the qubits, and each qubit must be addressed individually through long wires, the number of qubits that can be integrated on a single chip and tested is severely limited. Bringing control circuitry into the cryogenic environment has become a much-awaited solution - yet this comes with its own challenges and limitations. One of the main challenges in the design of complex cryogenic circuits is the lack of process design kits (PDKs) at low temperatures which, in turns, mirrors a lack of understanding and modelling of MOSFETs at liquid He temperature and below. Although in the last decade several works enlightened many critical aspects of cryogenic devices' static operation \cite{beckers_cryogenic_2018, bohuslavskyi_cryogenic_2019, catapano_zero_2023, ghibaudo_modelling_2020}, the understanding of defects dynamics and the resulting reliability issues requires more effort \cite{grill_reliability_2020, michl_evidence_2021}. In particular, low frequency noise (LFN) was shown to not scale linearly with temperature \cite{asanovski_understanding_2023, cardoso_paz_performance_2020, kiene_cryogenic_2024, oka_origin_2023}, as predicted by a variety of models that refer to carrier number fluctuations (CNFs) with correlated mobility fluctuations (MFs) \cite{ghibaudo_improved_1991, ghibaudo_electrical_2002}. This "excess 1/f noise" was proposed to be caused by density of states (DOS) bandtails \cite{asanovski_understanding_2023}, which are  shown to be responsible for the subthreshold slope saturation at cryogenic temperatures \cite{bohuslavskyi_cryogenic_2019, ghibaudo_modelling_2020, beckers_theoretical_2020}. However, understanding the origin of this excess noise is extremely challenging, since LFN is generally measured on large area devices and therefore it provides the average response of thousands of defects, making the characterization of individual traps impractical. For this reason, single defect spectroscopy techniques, such as random telegraph noise (RTN), were recently employed on small area devices to get insights on trap physics at cryogenic temperatures \cite{michl_evidence_2021, inaba_determining_2023, wang_new_2024}. In some studies, RTN was attributed to localized band-edge states \cite{inaba_determining_2023} or to interface defects \cite{michl_efficient_2021}, whereas in \cite{wang_new_2024} it was shown that RTN can also originate from defects far from the $\mathrm{Si-SiO_2}$ interface. In most RTN works, authors focused on the characterization of few selected devices, and the conclusions obtained can be hardly generalized to a statistically significant ensemble of defects. In other words, it is not possible to fully comprehend the statistical physical behaviour of a specific technology nor to link the activity of few defects with the unexpected 1/f behaviour of large area devices. To bridge this gap, we therefore use large-scale statistics of individual devices. We employ random telegraph noise characterization on an entire array of 2500 n-type MOSFETs. Using the ensemble statistics, we obtain distributions of single trap properties that can be linked to the physical mechanisms responsible for the 1/f noise across temperatures.\\
	 This work is structured as follows. First, RTN amplitudes are extracted using a state-of-the-art Machine Learning (ML) algorithm together with a Hidden Markov Model (HMM) algorithm. Afterwards, their distributions across temperatures and overdrives are plotted and extensively discussed. Threshold voltage shifts due to charged defects are exponentially distributed according to different modalities at different temperatures. By fitting these distributions, the average threshold voltage shift $\eta$ together with the statistical defect locations are extracted. Original time traces are then used to reconstruct the low frequency noise spectra of the entire array, which follows a 1/f behaviour down to cryogenic temperatures. The LFN is higher at $\mathrm{T=5}$ K, in excellent agreement with the observations already reported in literature \cite{asanovski_understanding_2023, cardoso_paz_performance_2020, oka_origin_2023}. The contribution to 1/f noise of defects coming from different modes and, therefore, different gate stack oxide layers, is further explored. Finally, the strong correlation between time constants and RTN amplitudes predicted by the elastic tunnelling model is tested \cite{nagumo_statistical_2010}.         
	
	\section{Results and Discussion}
	\subsection{Chip design and characterization}\label{sec2}
	
	\begin{figure}
		\centering
		\captionsetup{labelformat=empty}
		\subfloat[][\label{fig:Array}]
		{\includegraphics[width=.45\textwidth]{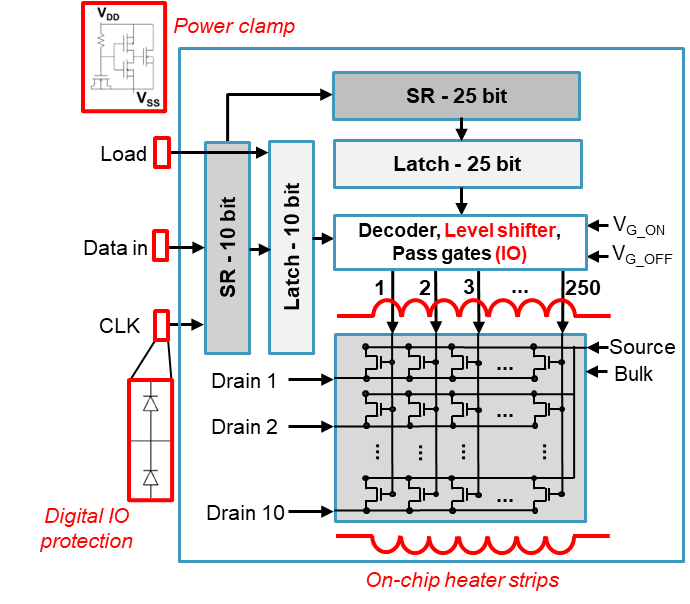}} \quad
		\subfloat[][\label{fig:IdVg}]
		{\includegraphics[width=.45\textwidth]{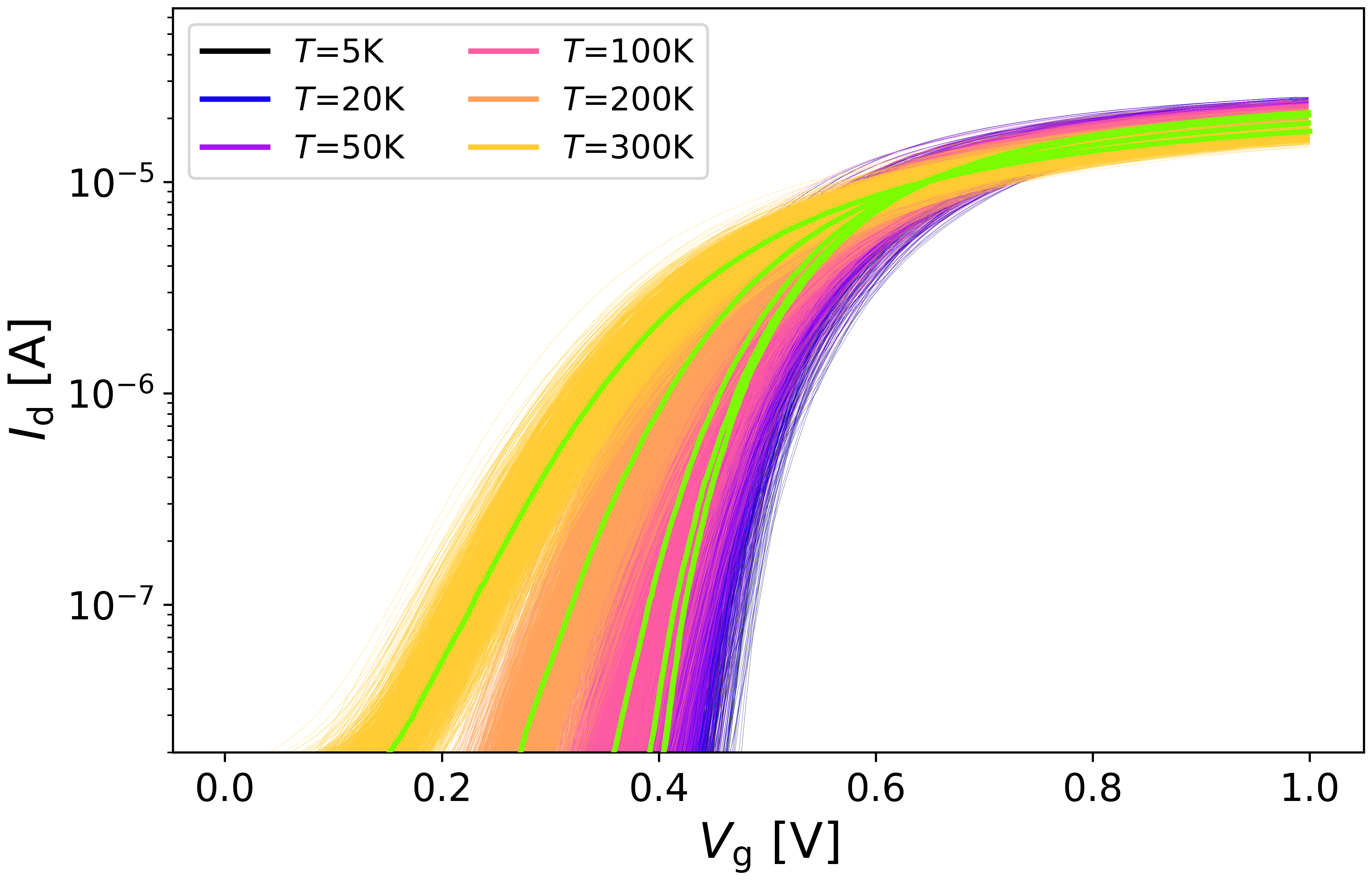}} \\
		\subfloat[][\label{fig:RTN}]
		{\includegraphics[width=.45\textwidth]{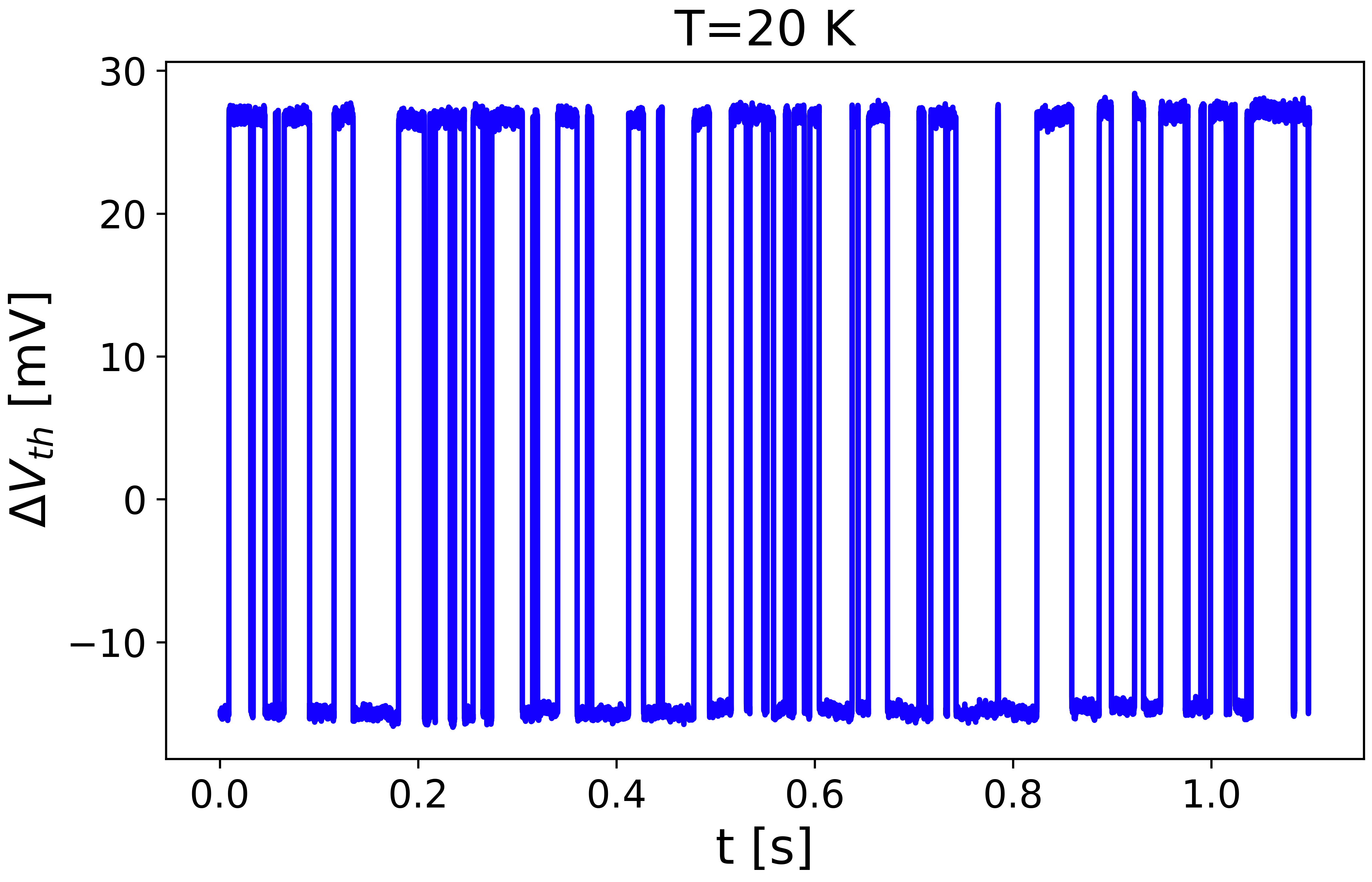}} \quad
		\subfloat[][\label{fig:RTN2}]
		{\includegraphics[width=.45\textwidth]{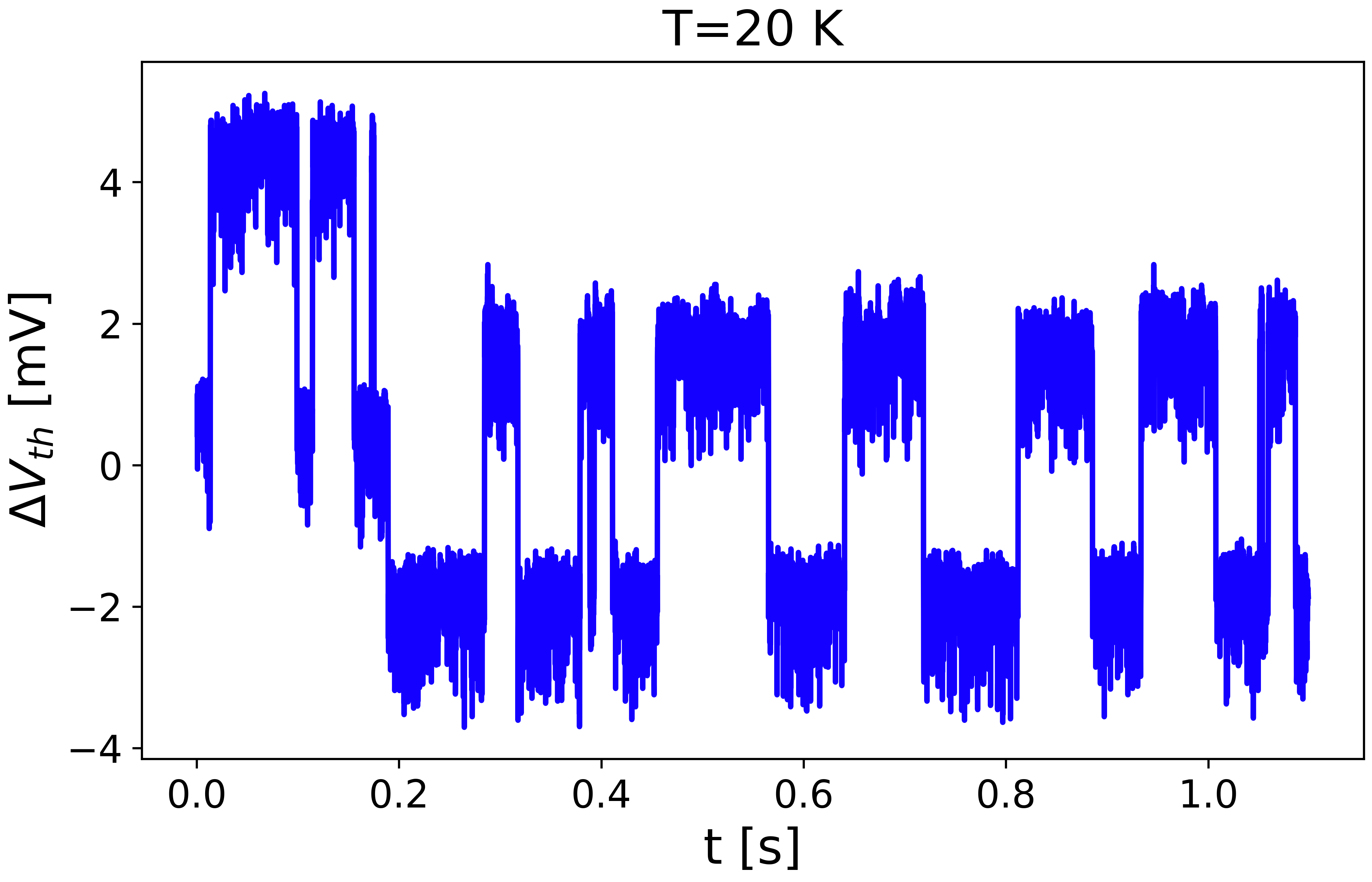}} \\
		\caption{\textbf{Fig.1} (a) The measured chip consists of 2500 individually addressable HKMG nMOS devices (WxL=90 nm x 28 nm). (b) Transfer characteristics $I_\mathrm{d}(V_\mathrm{gs})$ of all the devices from 300 K down to 5 K. (c) Drain current at stationary overdrive voltage showing a single defect, two levels RTN. (d) A similar measurement but showing two defects RTN.}
		\label{fig:Fig1}
	\end{figure}
	
	An array of 2500 nominally identical and minimally-spaced nanoscale $\mathrm{SiO_2/HfO_2}$ nMOS devices (WxL = 90 nm x 28 nm) was used in this work (\cref{fig:Array}). It consists of 250 individually-addressable gate lines, each connected to 10 devices. The structure was designed and fabricated using a commercial 28 nm high-k metal gate CMOS technology (see Methods). The chip was bonded on a custom designed PCB and placed in a Lakeshore CPX-LVT cryogenic probe station, equipped with a custom designed 48-lines break-out box \cite{grill_temperature_2022}. The measurements were performed on specifically designed hardware, which allowed to not only program and measure the array but also to change the sample temperature, enabling semi-automated measurement sequences.\\
	The array was tested at six different temperatures, from T=300 K down to T=5 K. At each temperature, the $I_\mathrm{d}(V_\mathrm{gs})$ characteristics in linear regime ($V_\mathrm{ds}=0.05$ V) of the entire device ensemble were collected (see \cref{fig:IdVg}). This step was used to remove the broken devices from the analysis, extracting the threshold voltage $V_\mathrm{th}$ of every working device and map the $I_\mathrm{d}(t)$ data to $\Delta V_\mathrm{th}$. The threshold voltage was calculated by using the Lambert-W function method \cite{karatsori_full_2015}, which was shown to be effective down to liquid helium temperature \cite{serra_di_santa_maria_lambert-w_2021}. Afterwards, the drain current $I_\mathrm{d}$ of each device was recorded for a total recording time of $t_r=1.1$ s, with a sampling time of $t_s=0.1$ ms, at five gate bias points: from $V_\mathrm{gs}=V_\mathrm{th}$ to $V_\mathrm{gs}=V_\mathrm{th}+0.2$ V, with a $V_\mathrm{gs}$ step $\Delta V_\mathrm{gs}=0.05$ V. The drain bias was fixed at $V_\mathrm{ds}=0.05$ V.\\
	In \cref{fig:Fig1} two examples of traces showing RTN are reported. In the first case (\cref{fig:RTN}), the RTN is produced by a single two-levels defect, which captures (low $\Delta V_\mathrm{th}$) and emits (high $\Delta V_\mathrm{th}$) individual carriers from the channel. The second case, instead, represents a more complicated signal, with two two-level defects active at the same time. A key aspect of this work is the development and the employment of an algorithm able to treat multiple defect signals. Indeed, this need turned out to be crucial for the RTN analysis at cryogenic temperatures, where the number of traces containing two defects was around 30\% of the total.\\
	To extract the defect average step height $\Delta V_\mathrm{th}$, as well as the average emission and capture time constants $\tau_{e,c}$, a method based on a physics-informed machine learning (ML) algorithm \cite{varanasi_physics-informed_2024} \cite{varanasi_rtninja_2025} coupled with a hidden Markov model (HMM) \cite{grill_charge_nodate} was developed. The ML algorithm allowed to automatically detect RTN in the $\Delta V_\mathrm{th}(t)$ traces, distinguishing between one or multi-defect signals. In case of single-defect RTN, $\Delta V_\mathrm{th}$ and $\tau_{e,c}$ were directly extracted from the signal reconstructed by the ML algorithm, whereas, in case of two active defects, the raw data were further analysed using the HMM. The benefits of this method are manifold: it requires little manual interaction, so it is suited for the treatment of large amount of data; given its complexity, it is reasonably fast (it can run on a personal computer); it allows the analysis of multi-defects RTN, as previously discussed.

	\subsection{Defects' step heights distributions}\label{sec3}
	
	\begin{figure}
		\centering
		\captionsetup{labelformat=empty}
		\subfloat[][\label{fig:CCDF 300K}]
		{\includegraphics[width=.45\textwidth]{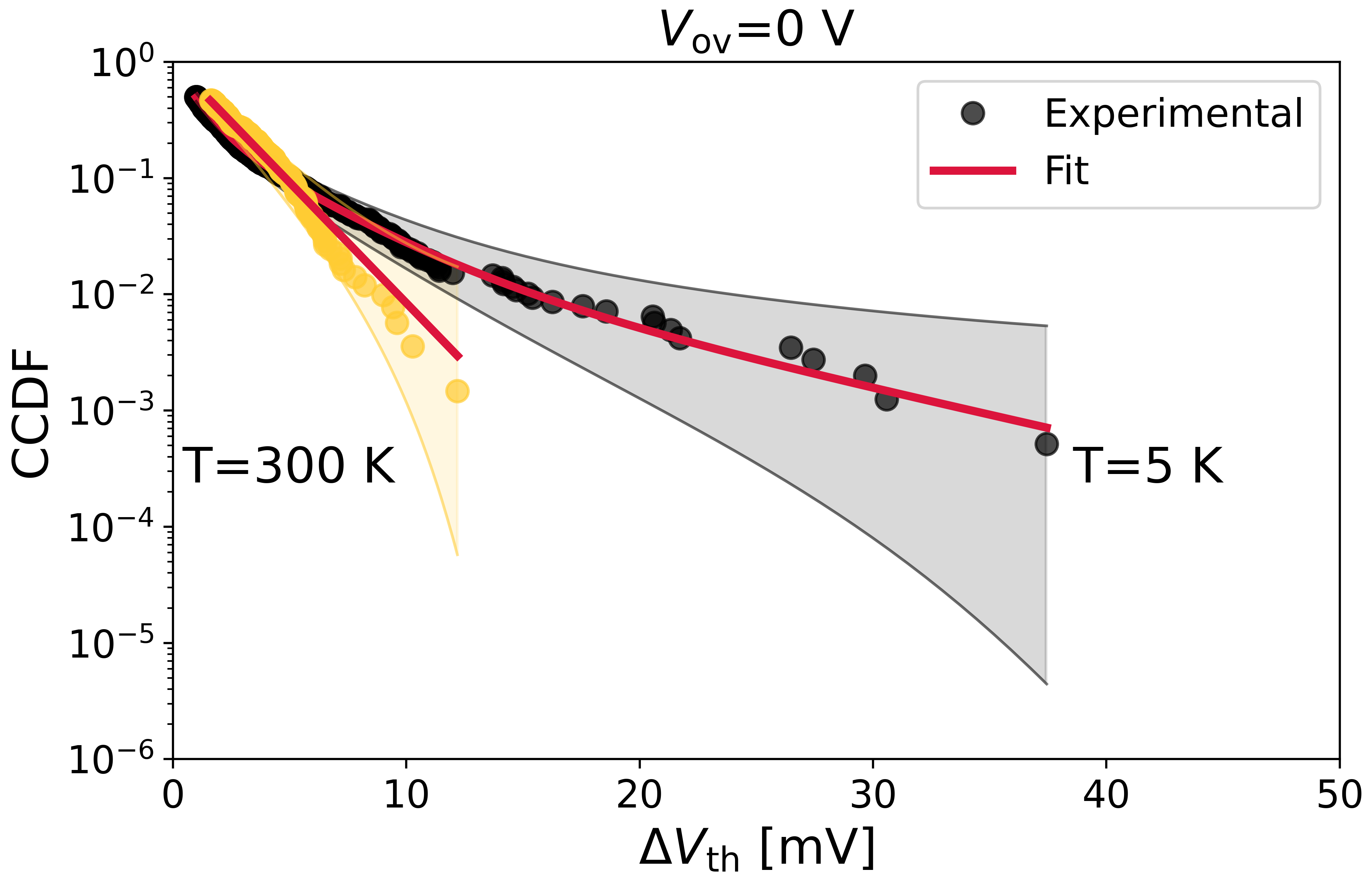}} \quad
		\subfloat[][\label{fig:PDF}]
		{\includegraphics[width=.45\textwidth]{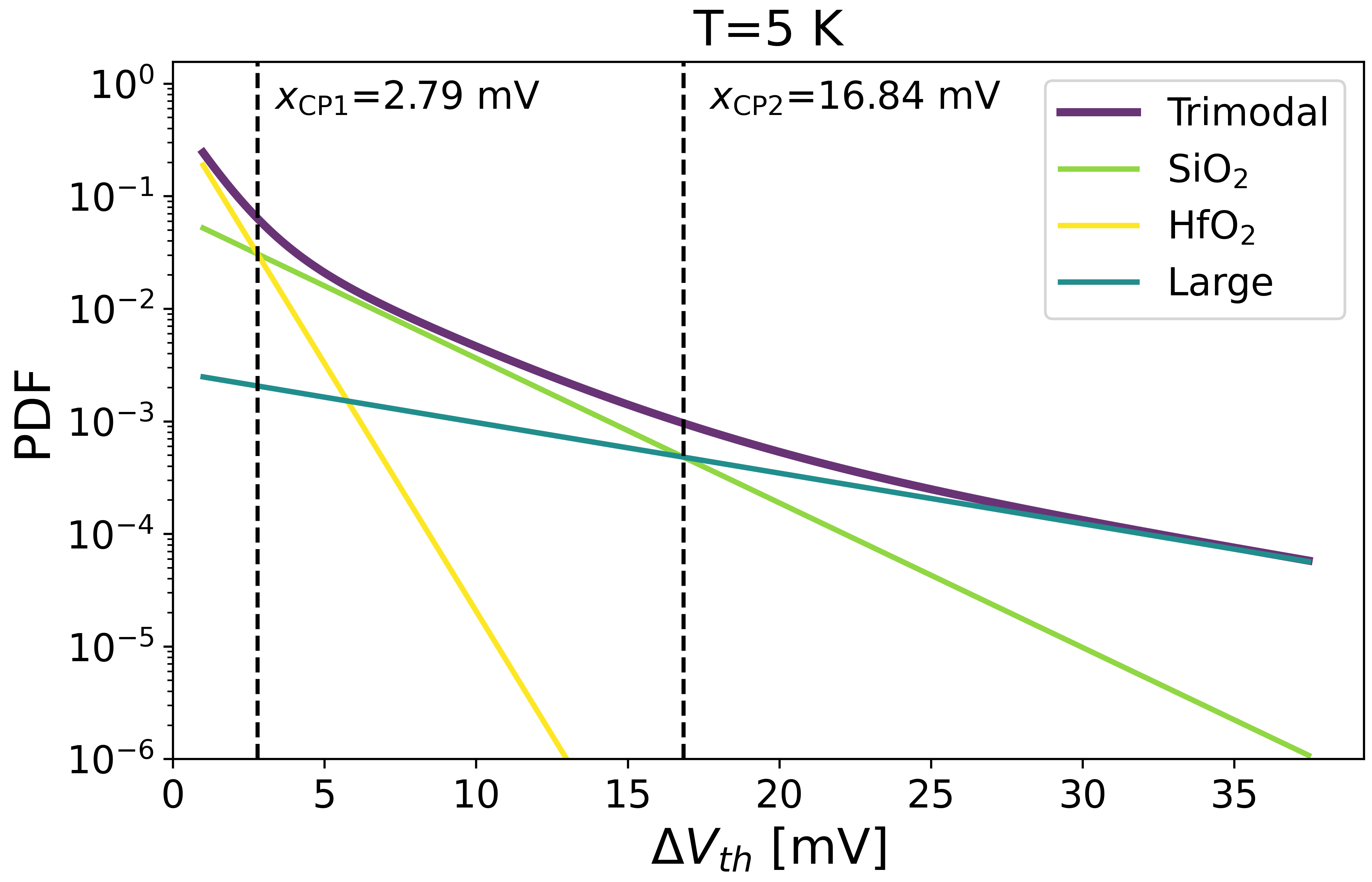}} \\
		\subfloat[][\label{fig:Prob}]
		{\includegraphics[width=.45\textwidth]{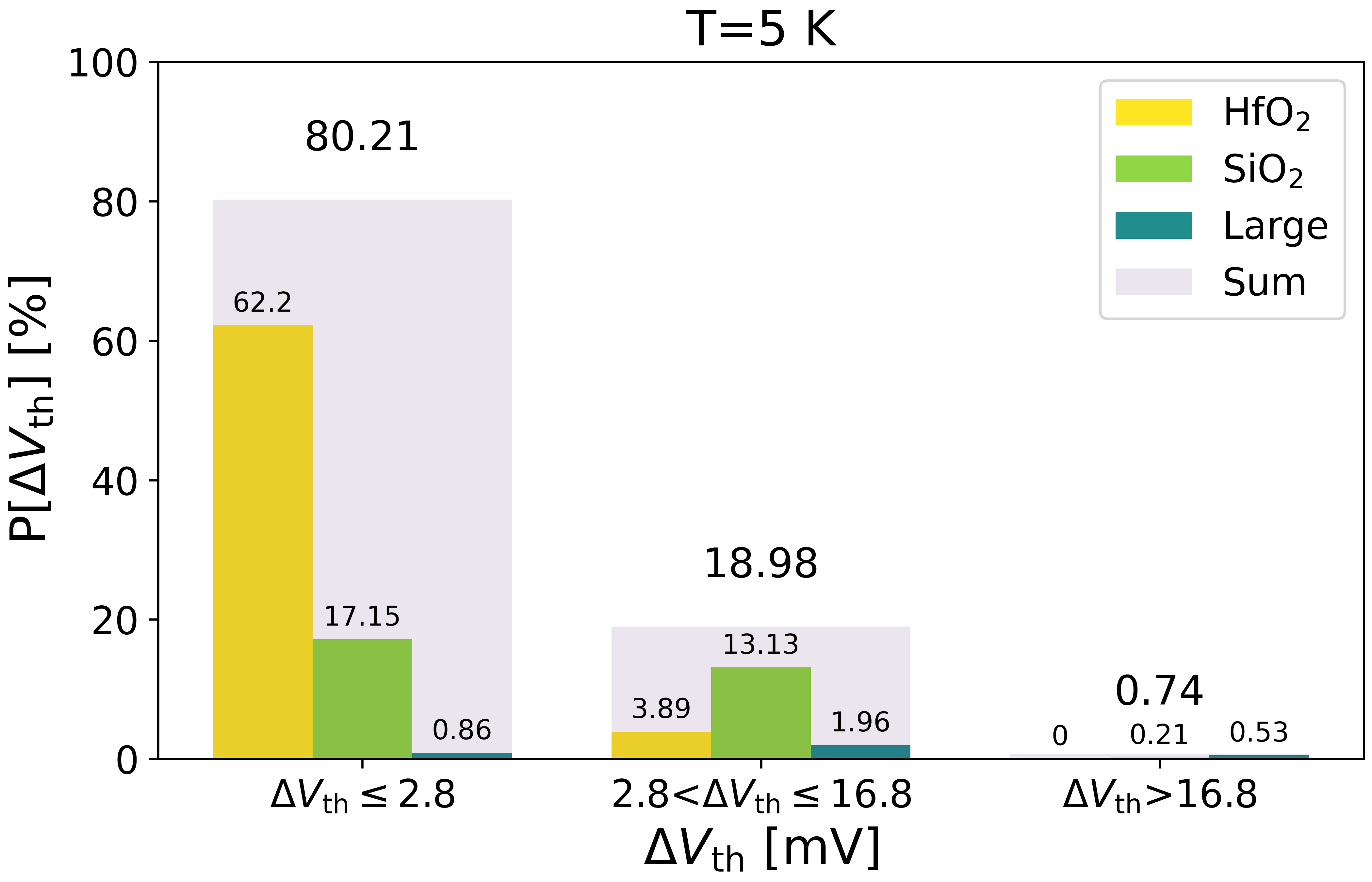}} \quad
		\subfloat[][\label{fig:eta}]
		{\includegraphics[width=.45\textwidth]{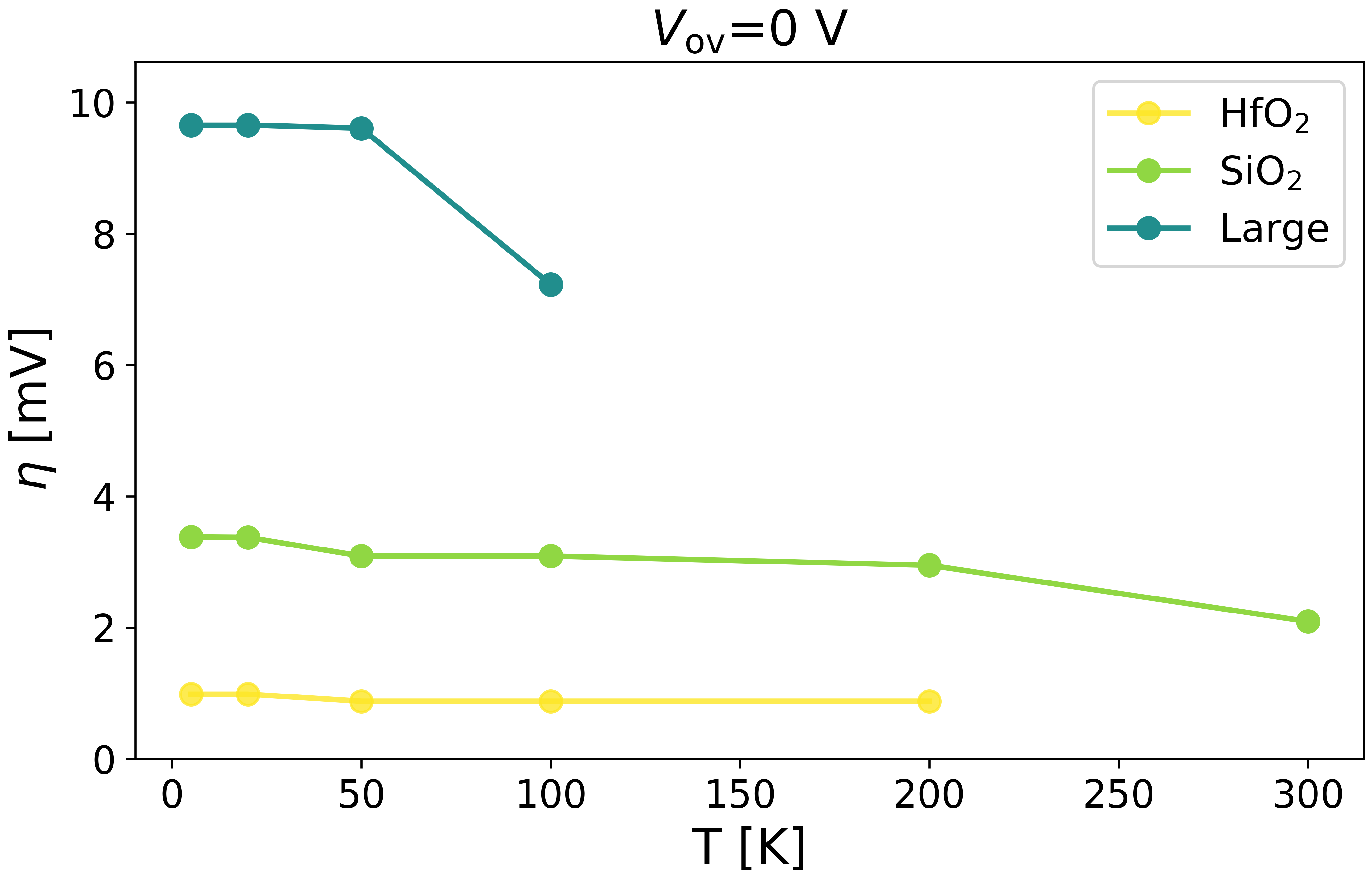}} \\
		\caption{\textbf{Fig.2} (a) At T=300 K, the $\Delta V_\mathrm{th}$ CCDF follows a monomodal exponential distribution, even though the oxide gate stack consists of $\mathrm{HfO_2/SiO_2}$. Small $\mathrm{HfO_2}$ defects are probably hidden in the background noise. At T=5 K, two additional modes appear. The smallest mode is now visible because of the smaller background noise and the higher measurement sensitivity due to the larger $g_m$. The large defects belonging to the third mode are likely caused by the higher sensitivity to disorder at cryogenic temperatures. (b) PDF of the components of the trimodal exponential $\Delta V_\mathrm{th}$ distribution at 5K a $V_\mathrm{ov}=0$ V. Each mode consists of an exponential distribution with given $\eta$ and $n$. The first mode is attributed to defects belonging to $\mathrm{HfO_2}$; the second mode is attributed to defects belonging to $\mathrm{SiO_2}$; the third mode is attributed to defects abnormally large. (c) Probability of a defect falling within particular range of values, defined by the calculated crossing points $x_\mathrm{CP1}$, $x_\mathrm{CP2}$. More than 80\% of detected defects have $\Delta V_\mathrm{th} \leqslant 2.8$ mV, and more than 60\% of defects within this interval belongs to $\mathrm{HfO_2}$, showing that bulk defects activity is still predominant at T=5 K. (d) $\eta (T)$ extracted from fitting the CCDF at different T. Its increase at low temperatures and at $V_\mathrm{ov}=0$ V is attributed to a larger sensitivity to disorder at cryogenic temperatures.}
		\label{fig:Fig2}
	\end{figure}
	
	 Random telegraph noise $V_\mathrm{th}$ fluctuations have been repeatedly reported to be exponentially distributed \cite{ghetti_comprehensive_2009, stampfer_semi-automated_2020, asenov_advanced_2008}, as well as the $\Delta V_\mathrm{th}$ steps extracted from BTI measurements \cite{kaczer_origin_2010, toledano-luque_degradation_nodate, toledano-luque_defect-centric_2012, toledano-luque_temperature_2011, waltl_separation_2020, tselios_evaluation_2022}. The complementary cumulative distribution function ($\mathrm{1 - CDF}$ or $\mathrm{CCDF}$) is usually employed in reliability studies. In the case of a monomodal exponential distribution, it is expressed as:
	 
	 \begin{equation}
	 	\label{eqn:exp}
	 	\mathrm{CCDF}=n\cdot \mathrm{exp}\Bigg(-\frac{\Delta V_\mathrm{th}}{\eta} \Bigg)
	 \end{equation}
	 \\
	 where $n$ is the total number of active defects, $\Delta V_\mathrm{th}$ is the threshold voltage shift due to a single charged defect and $\eta$ is the mean threshold voltage shift. Percolation theory \cite{ghetti_comprehensive_2009, takeuchi_single-charge-based_nodate} proved that the exponential distribution of $\Delta V_\mathrm{th}$ is caused by non uniformities in the surface potential and, in turn, in the drain current. Indeed, at weak and moderate inversion, the channel is not fully formed and carriers flow through paths with different conductivities. These inhomogeneities were generally attributed to random dopants fluctuations (RDF) in the $\mathrm{Si}$ substrate \cite{ghetti_comprehensive_2009, tselios_evaluation_2022}, but it has been shown that oxide defects can contribute as well \cite{toledano-luque_degradation_nodate}. Furthermore, depending on the gate stack layer, distributions can have different modalities. Technologies with a single gate oxide stack often show the monomodal CCDF defined in \text{eq.}\eqref{eqn:exp}, since all the defects probed belong to the same band and, in turn, to the same oxide \cite{tselios_evaluation_2022}. The case of multi-layers oxide stacks is discussed later in the text.\\
	 It is worth mentioning that, by assuming an homogeneous current flow in the transistor channel, RTN amplitudes should be uniformly distributed, with a threshold voltage shift given by
	 
	\begin{equation}
		\label{eqn:CSA}
		\Delta V_\mathrm{th}=-\frac{q}{\epsilon_0 \epsilon_r WL}t_{ox}\Big(1-\frac{x_\mathrm{T}}{t_{ox}} \Big)
	\end{equation}
	\\
	where $q$ is the elementary charge, $\epsilon_0$ is the vacuum permittivity, $\epsilon_r$ is the relative permittivity of the material, $t_\mathrm{ox}$ is the oxide thickness and $x_\mathrm{T}$ is the trap position with respect to the $\mathrm{Si-SiO_2}$ interface. \text{Eq.}\eqref{eqn:CSA} is usually referred to as "charge sheet approximation (CSA)". CSA assumes the oxide charge to be uniformly distributed over a fictitious sheet in the insulator. However, neglecting the impact of percolative conduction due to RDF causes a severely underestimation of the average impact of defects $\eta$. Furthermore, CSA cannot explain the large $\Delta V_\mathrm{th}$ shifts experimentally observed in the tail of the distribution \cite{kaczer_origin_2010}. \\
	
	\noindent In \cref{fig:CCDF 300K} are plotted the CCDFs of the step heights extracted from the measurement data of approximately 2500 devices at both $\mathrm{T=300}$ K and $\mathrm{T=5}$ K, and $V_\mathrm{ov}=0$ V. With decreasing temperature, the shape of the CCDF changes. To determine the modality of each exponential distribution in a statistically rigorous manner, a fitting method based on the maximization of the likelihood function was developed together with an outliers detection algorithm (see Methods section). Furthermore, this fitting procedure allows the extraction of $n$ and $\eta$ (see \text{eq.}\eqref{eqn:exp}). It is worth pointing out that $n$ includes also the undetected defects (more details can be found in the Methods section). At $\mathrm{T=300}$ K the CCDF is monomodal, although the technology used in this work consists of devices with a $\mathrm{SiO_2/HfO_2}$ stack. Indeed, in case of multi-layers stacks, single-defects step heights generally follow a bimodal distribution: the low-$\eta$ mode belongs to $\mathrm{HfO_2}$ bulk defects, whereas the high-$\eta$ mode is attributed to $\mathrm{SiO_2}$ interface defects \cite{toledano-luque_temperature_2011, weckx_defect-centric_2015}. This division is based on electrostatic considerations: $\mathrm{HfO_2}$ defects are smaller compared to $\mathrm{SiO_2}$ ones, since they are more far away from the channel and their electrostatic impact is lower. In the present case, most of $\mathrm{HfO_2}$ defects probably have a $\Delta V_\mathrm{th}$ smaller than the detection limit of both the ML algorithm and the HMM used in the time traces analysis. At $\mathrm{T=300}$ K and $V_\mathrm{ov}=0$ V, the detection limit was set to $\Delta V_\mathrm{th,min}=1.6$ mV. It is worth pointing out that this argument finds confirmation in \cref{fig:eta}, where the extracted values of $\eta$ are plotted. Even at $\mathrm{T=5}$ K, where $\eta$ is generally found to be higher, as it will be discussed later, $\eta_\mathrm{HfO_2} \approx 1$ mV. Therefore, the CCDF shown in \cref{fig:CCDF 300K} belongs only to $\mathrm{SiO_2}$ defects. At this point, it can be instructive to compare the experimental $\eta \equiv \eta_{SiO_2}$ with $\eta_{0,SiO_2}$ predicted by the CSA. Coherently with the literature \cite{kaczer_origin_2010, toledano-luque_degradation_nodate, tselios_evaluation_2022}, $\eta_{SiO_2}=2.1$ mV is more than two times larger than $\eta_{0,SiO_2}=0.89$ mV. As the temperature drops to $\mathrm{T=200}$ K, $\mathrm{HfO_2}$ defects start to enter the detection window, because of both a reduction of the background thermal noise and an increase of the transistors' $g_\mathrm{m}$. Indeed, since $\Delta I_\mathrm{d}=\Delta V_\mathrm{th} \cdot g_\mathrm{m}$, small threshold voltage shifts are easier to detect at low temperatures because they result in a larger $\Delta I_\mathrm{d}$. This results in the formation of a low-$\eta$ branch in the CCDF, which is bimodal at $\mathrm{200}$ K (see Supplement Material).\\
	
	\noindent Starting from $\mathrm{T=100}$ K (see Supplement Material), and down to $\mathrm{T=5}$ K (\cref{fig:CCDF 300K}), $\Delta V_\mathrm{th}$ distributions change modality again, with the appearance of another branch at very large $\Delta V_\mathrm{th}$s. The three modes are highlighted in \cref{fig:PDF}, where the probability density function (PDF) of the trimodal exponential fit is plotted. This latter is given by the sum of three independent monomodal exponential PDFs, each one being described by a different set of $n$ and $\eta$. The first mode is attributed to $\mathrm{HfO_2}$ defects. The second mode is linked to $\mathrm{SiO_2}$ defects, which enables a direct investigation of their temperature scaling from $\mathrm{T=5}$ K up to $\mathrm{T=300}$ K, as discussed later. The nature and origin of the third, large $\Delta V_\mathrm{th}$ mode will be deeply investigated later in this section. \\
	
	\noindent A defect with a given amplitude $\Delta V_\mathrm{th}$ is associated with each mode with a certain probability. In \cref{fig:PDF}, there are two $\Delta V_\mathrm{th}$ values that have the same probability to belong to two modes: these are given by the intersection points $x_\mathrm{CP1}$ and $x_\mathrm{CP2}$. A threshold voltage shift due to charged defects lying on the left side of $x_\mathrm{CP1}$, for example, is more likely to fall in the $\mathrm{HfO_2}$ branch. The probability that the amplitude of a detected oxide defect falls into intervals of values defined by $x_\mathrm{CP1}$ and $x_\mathrm{CP2}$ is calculated using the cumulative distribution function (CDF) and it is plotted in \cref{fig:Prob}. More than 80\% of measured defects have $\Delta V_\mathrm{th} \leqslant 2.8$ mV, and more than 60\% of these belongs to $\mathrm{HfO_2}$. Therefore, contrary to what has been previously published on RTN at cryogenic temperatures \cite{oka_origin_2023} \cite{inaba_determining_2023}, most of the activity seems to come from bulk defects, which do not freeze out. In \cref{fig:eta}, the temperature dependence of $\eta$ at $V_\mathrm{ov}=0$ V is plotted. Globally, $\eta$ increases at cryogenic temperatures, independently on the branch considered. Defects linked to the third mode, however, appear to be more sensitive to temperature scaling. $\mathrm{SiO_2}$ and $\mathrm{HfO_2}$ defects also show an increase in their average threshold voltage shift as the temperature is lowered. As it will be discussed below, this trend of $\eta(\mathrm{T})$ can be effectively interpreted in the framework of percolation theory.\\
	
\begin{figure}
		\centering
		\captionsetup{labelformat=empty}
		\subfloat[][\label{fig:Percolation1}]
		{\includegraphics[width=.45\textwidth]{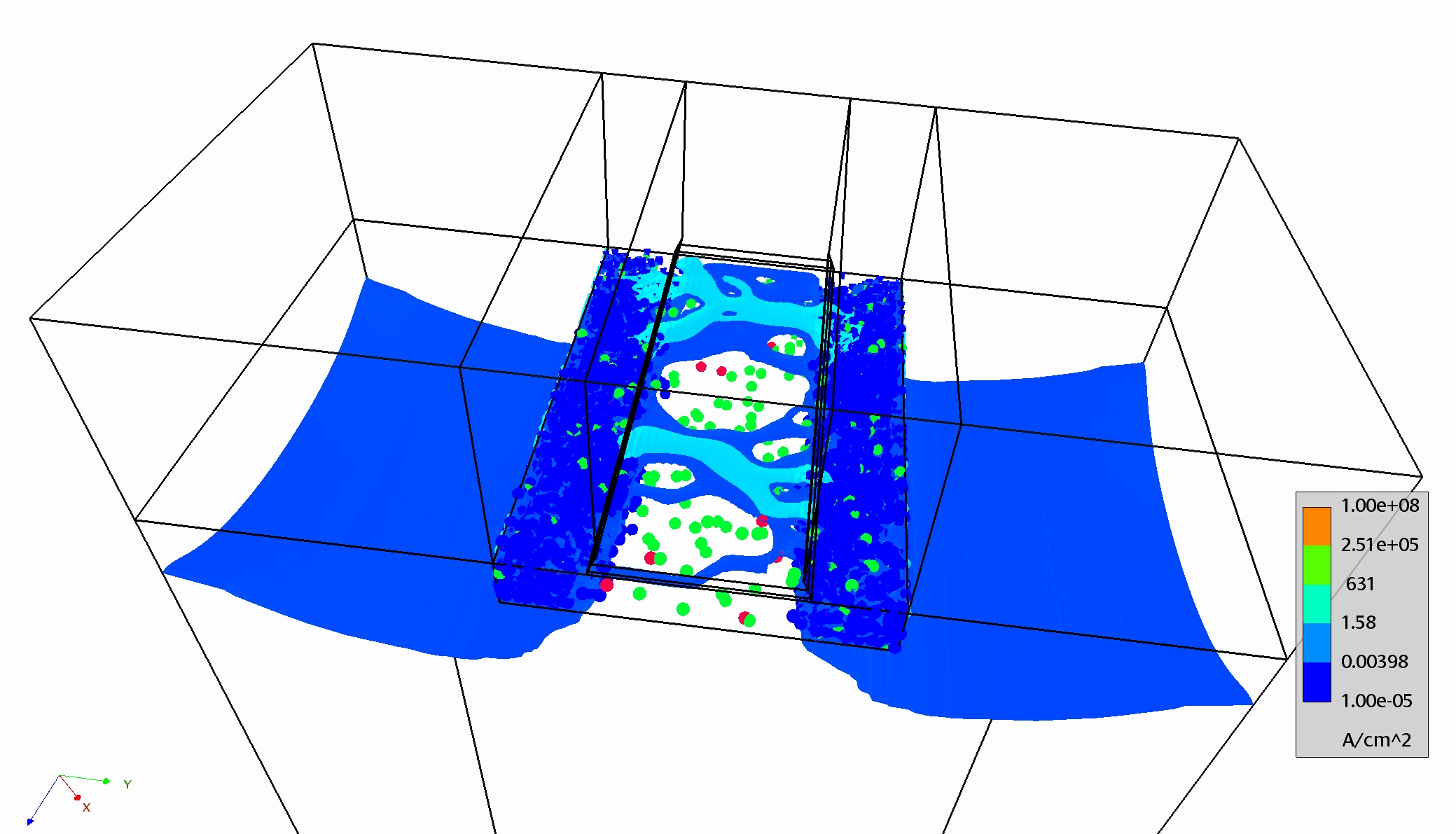}} \quad
		\subfloat[][\label{fig:Percolation2}]
		{\includegraphics[width=.45\textwidth]{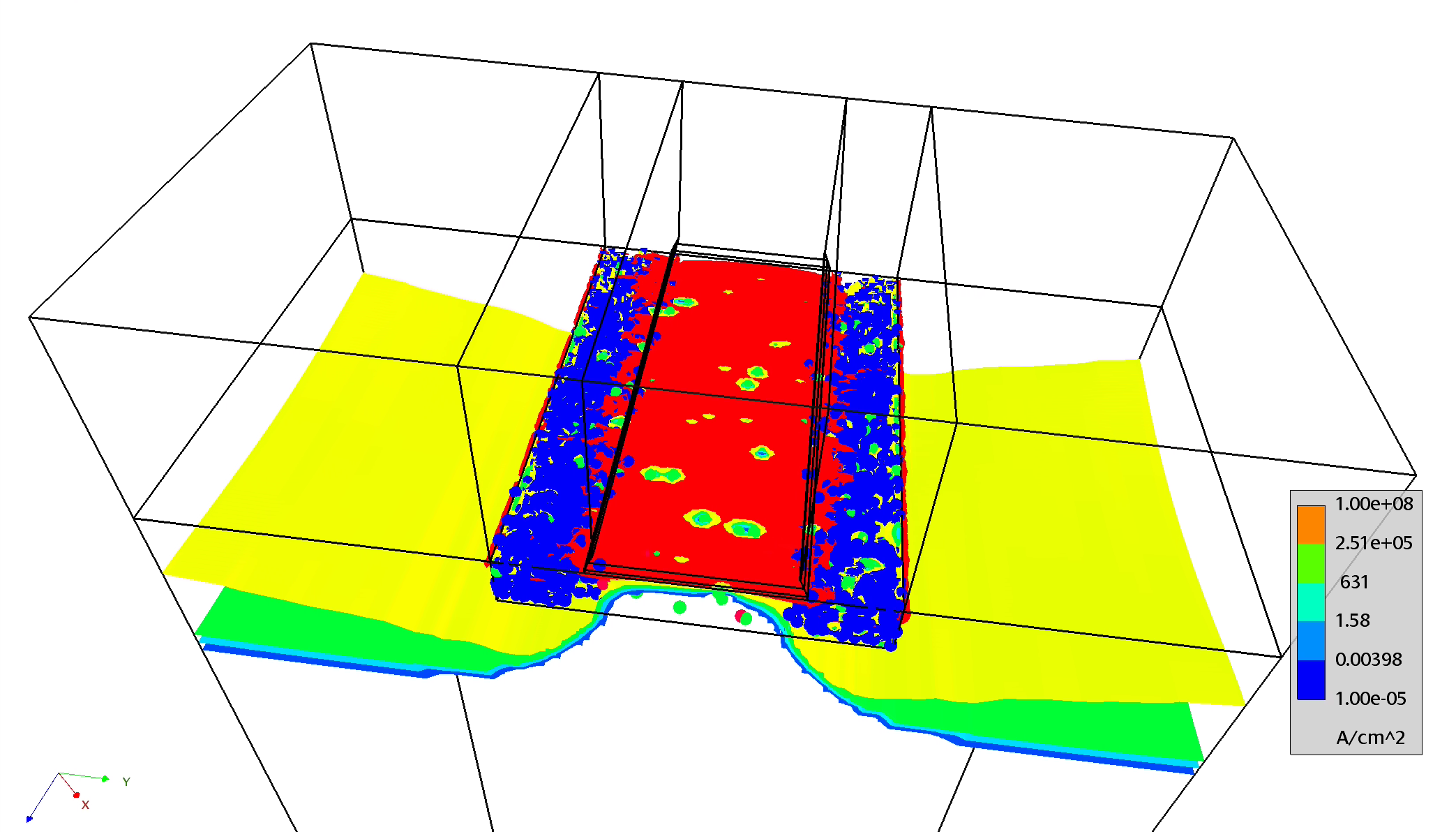}} \\
		\subfloat[][\label{fig:CCDF 5K}]
		{\includegraphics[width=.45\textwidth]{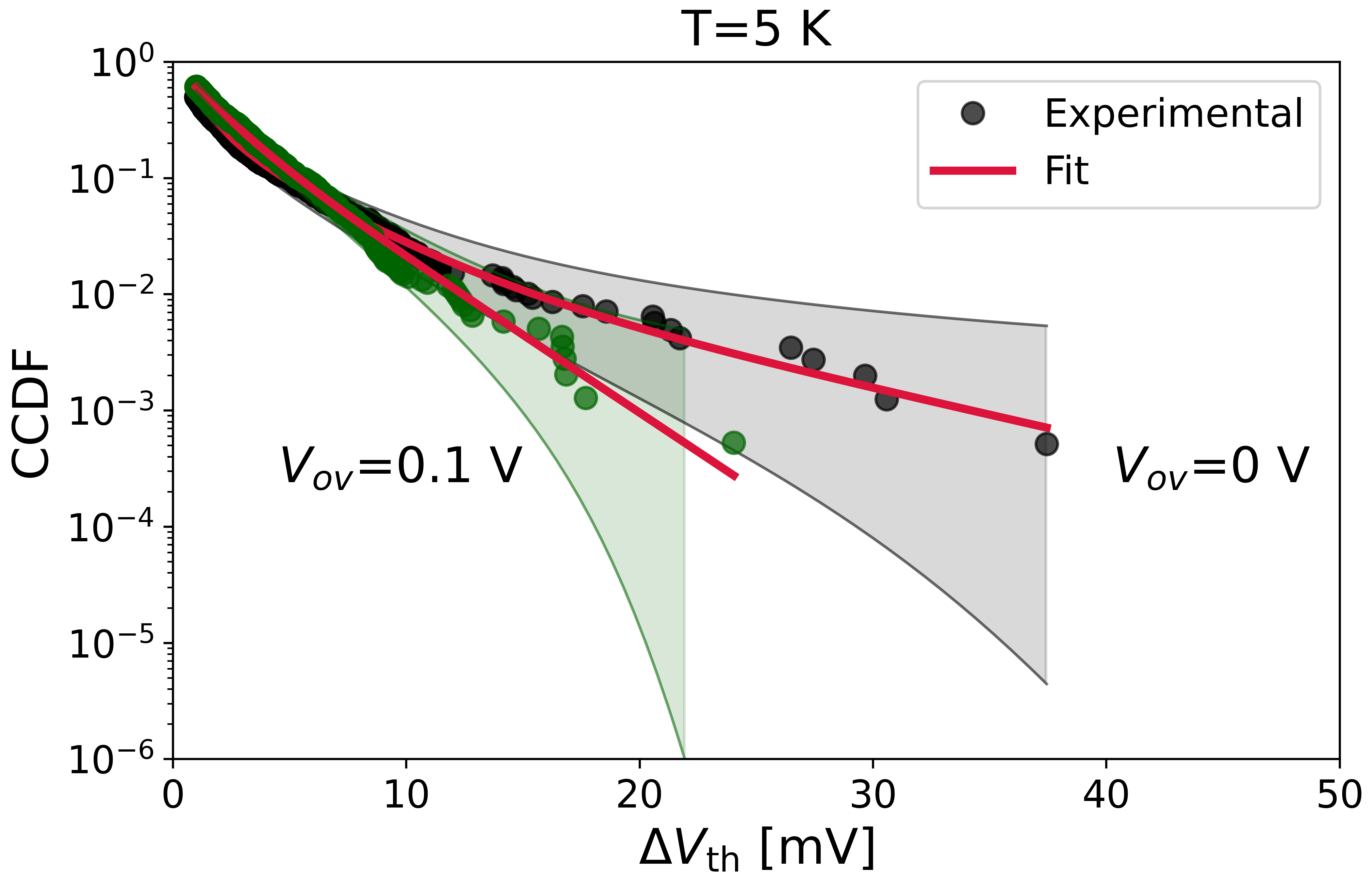}} \quad
		\subfloat[][\label{fig:eta_SiO2}]
		{\includegraphics[width=.45\textwidth]{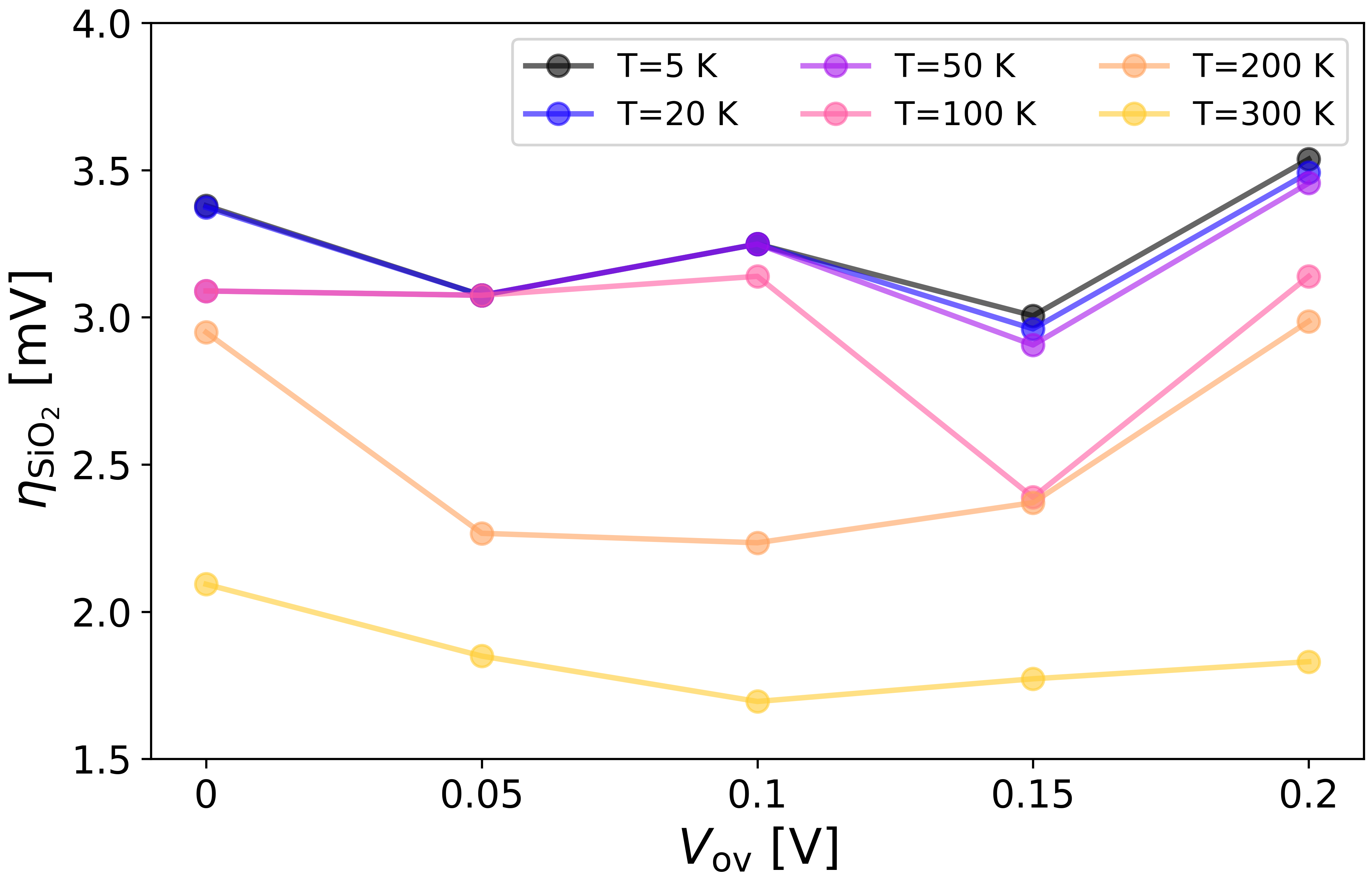}} \\
		\caption{\textbf{Fig.3}  (a) Drain current density simulated with a commercial TCAD tool at $\mathrm{T=4.2}$ K. At $V_\mathrm{gs}<V_\mathrm{th}$, the drain current is not homogeneous due to the presence of both random dopants and oxide defects. This non homogeneity translates in the formation of conductive paths, where carriers can flow through, and insulating regions, where carriers are forbidden to flow. Percolation generally increases at cryogenic temperatures, since particles' kinetic energy decreases and surface potential fluctuations can provoke the formation of localized states which, in turns, are responsible for the anomalously large $\Delta V_\mathrm{th}$ observed. (b) By increasing $V_\mathrm{ov}$, the channel gets stronger and, therefore, more homogeneous, and percolation is reduced. (c) At higher $V_\mathrm{ov}$, the third mode disappears from the CCDF, since the electrostatic impact of defects is reduced by the stronger channel. (d) $\eta_\mathrm{SiO_2}$ is rather constant as function of $V_\mathrm{ov}$, independently on the temperature.}
		\label{fig:Fig3}
	\end{figure}

	\noindent In order to interpret the shape of the CCDF at cryogenic temperatures (see \cref{fig:CCDF 300K} and \cref{fig:PDF}), it is worth mentioning that, in the field of cryogenic electronics, percolation theory was proposed in the late '70s to interpret the smooth change from metallic-like to thermally-activated transport in MOSFET's inversion layer when it operates around and below $V_\mathrm{th}$ and at $\mathrm{T=4.2}$ K \cite{arnold_disorder-induced_1974, ghibaudo_transport_1986, catapano_cryogenic_2023}. This metal-insulator transition \cite{mott_anderson_1975} is generally attributed to surface potential fluctuations caused by $\mathrm{Si-SiO_2}$ interface disorder. Macroscopically, these fluctuations are usually modelled as an additional density of states below the conduction (valence) band edge \cite{bohuslavskyi_cryogenic_2019, ghibaudo_modelling_2020, beckers_theoretical_2020}. Therefore, percolation theory has been employed independently to describe two apparently different phenomena, both originating from surface potential fluctuations: electrical conduction in band tail states at cryogenic temperatures and statistical distribution of threshold voltage shifts due to charged defects at room temperature. Starting from these considerations, percolation theory is here proposed to interpret the $\Delta V_\mathrm{th}$ distributions at $\mathrm{T \leqslant 100}$ K (see \cref{fig:CCDF 300K} and Supplement Materials). In this framework, the observed CCDFs are a manifestation of both the increased sensitivity to disorder and the appearance of interface localized states. 
	Fluctuations in surface potential mirrors in a non uniform distribution of inversion carriers along the channel. If they have enough thermal energy, the main consequence of such fluctuations is a non homogeneous current flow from source to drain which, as previously mentioned, is responsible for the exponential distribution of $\Delta V_\mathrm{th}$s. As temperature drops, carriers thermal energy decreases accordingly, and potential fluctuations can give rise to regions where carriers are energetically prevented to flow. It is crucial to highlight that, at room temperature, the channel is modulated by disorder but it is continuous. At cryogenic temperatures, instead, the channel is a mixture of conductive paths and insulating regions \cite{arnold_disorder-induced_1974}. In \cref{fig:Percolation1}, it is plotted a qualitative TCAD simulation performed at $\mathrm{T=4.2}$ K to illustrate this phenomenon: at low $V_\mathrm{gs}$ the channel is not fully formed and conductive paths coexist together with insulating areas. The increased channel inhomogeneity amplifies the electrostatic impact of all defects at low temperatures, in agreement with the $\eta(\mathrm{T})$ trend previously discussed and plotted in \cref{fig:eta}. RTN fluctuations producing the third branch in \cref{fig:CCDF 300K} are therefore interpreted as the result of the activity of those defects whose electrostatic impact is particularly enhanced by the larger channel inhomogeneity. On the other hand, especially around $\mathrm{T=5}$ K, potential fluctuations can give rise to localized states, which contribute to transport through thermally-activated and also tunnelling processes. Some of the large $\Delta V_\mathrm{th}$ in \cref{fig:CCDF 300K} could be caused by the activity of these localized states as well, but further analyses are required to confirm that.\\

	\noindent Increasing $V_\mathrm{ov}$, the modality of the CCDF changes, passing from trimodal to bimodal, as shown in \cref{fig:CCDF 5K}. The third branch in the CCDF vanishes, since the channel gets stronger and percolation transport is mitigated, as qualitatively illustrated in \cref{fig:Percolation2}. On the other hand,  the small $\mathrm{HfO_2}$ mode is still present. A further increase in $V_\mathrm{ov}$ wipes out also the $\mathrm{HfO_2}$ branch and, at high overdrives, only the $\mathrm{SiO_2}$ defects are visible (see Supplement Material). This can be attributed to both a reduction in measurement sensitivity and to a shift in the active energy area scanned. Concerning the first argument, at higher $V_\mathrm{ov}$ it is harder to detect small defects, since $I_\mathrm{d}$ is larger. Moreover, changing the gate bias, the Fermi level changes as well, and the energy area of the defect bands swept changes accordingly. Therefore, at $V_\mathrm{ov} > 0.1$ V, CCDFs are attributed only to the activity of defects in the $\mathrm{SiO_2}$ layer.  \\
	      
	 \noindent In \cref{fig:eta_SiO2}, $\eta_\mathrm{SiO_2}$ is plotted as function of $V_\mathrm{ov}$. Globally, at lower temperatures $\eta_\mathrm{SiO_2}$ is higher, independently on $V_\mathrm{ov}$, as already shown in \cref{fig:eta}. This is consequence of both the larger percolation and the shift in the active energy area scanned by the Fermi level at cryogenic temperatures. On the other hand, $\eta_\mathrm{SiO_2}$ is almost independent on $V_\mathrm{ov}$.  
	
	\subsection{Low frequency noise reconstruction}\label{sec4}
	
	\begin{figure}
		\centering
		\captionsetup{labelformat=empty}
		\subfloat[][\label{fig:Np}]
		{\includegraphics[width=.45\textwidth]{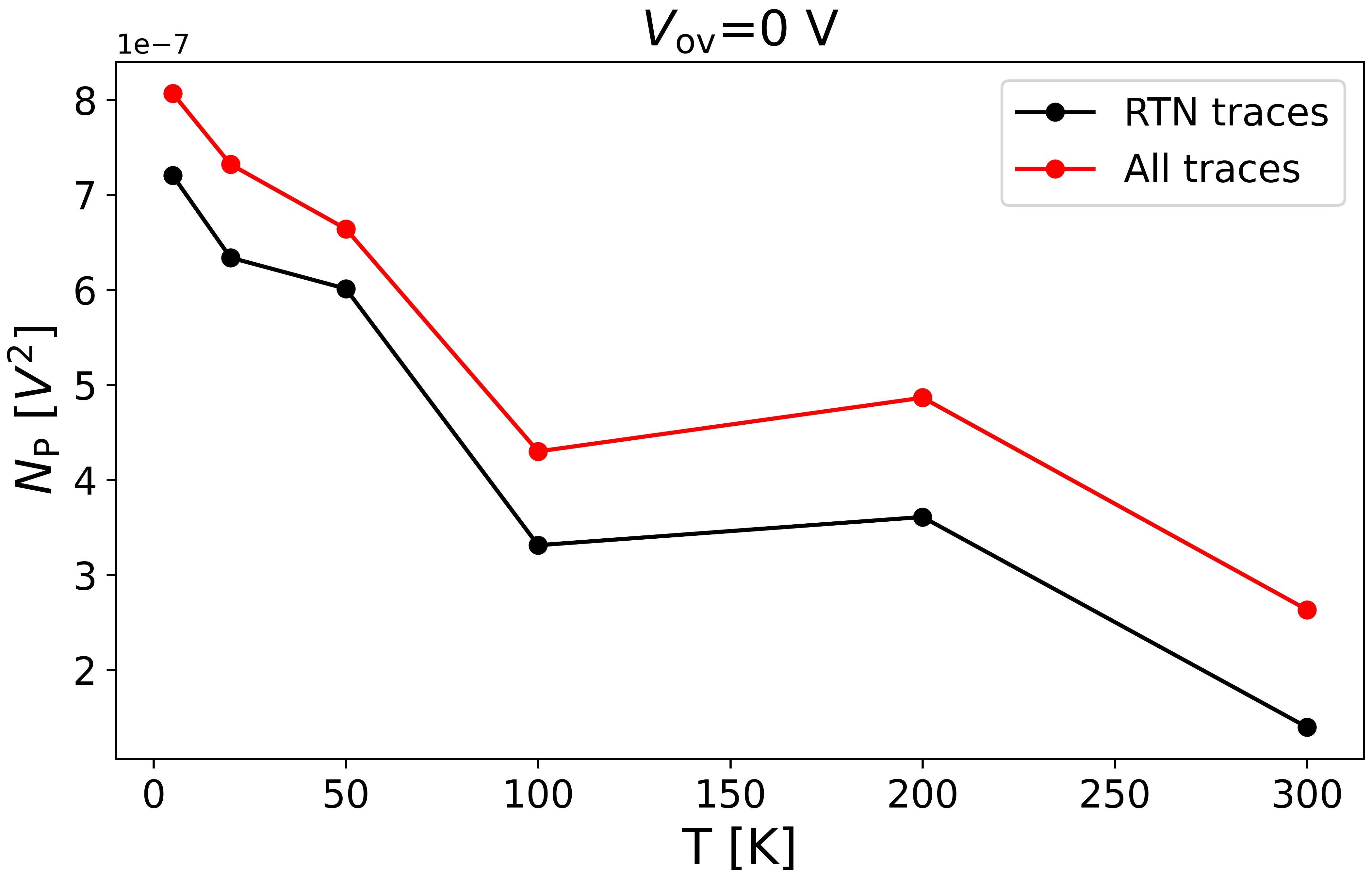}} \quad
		\subfloat[][\label{fig:SVg}]
		{\includegraphics[width=.45\textwidth]{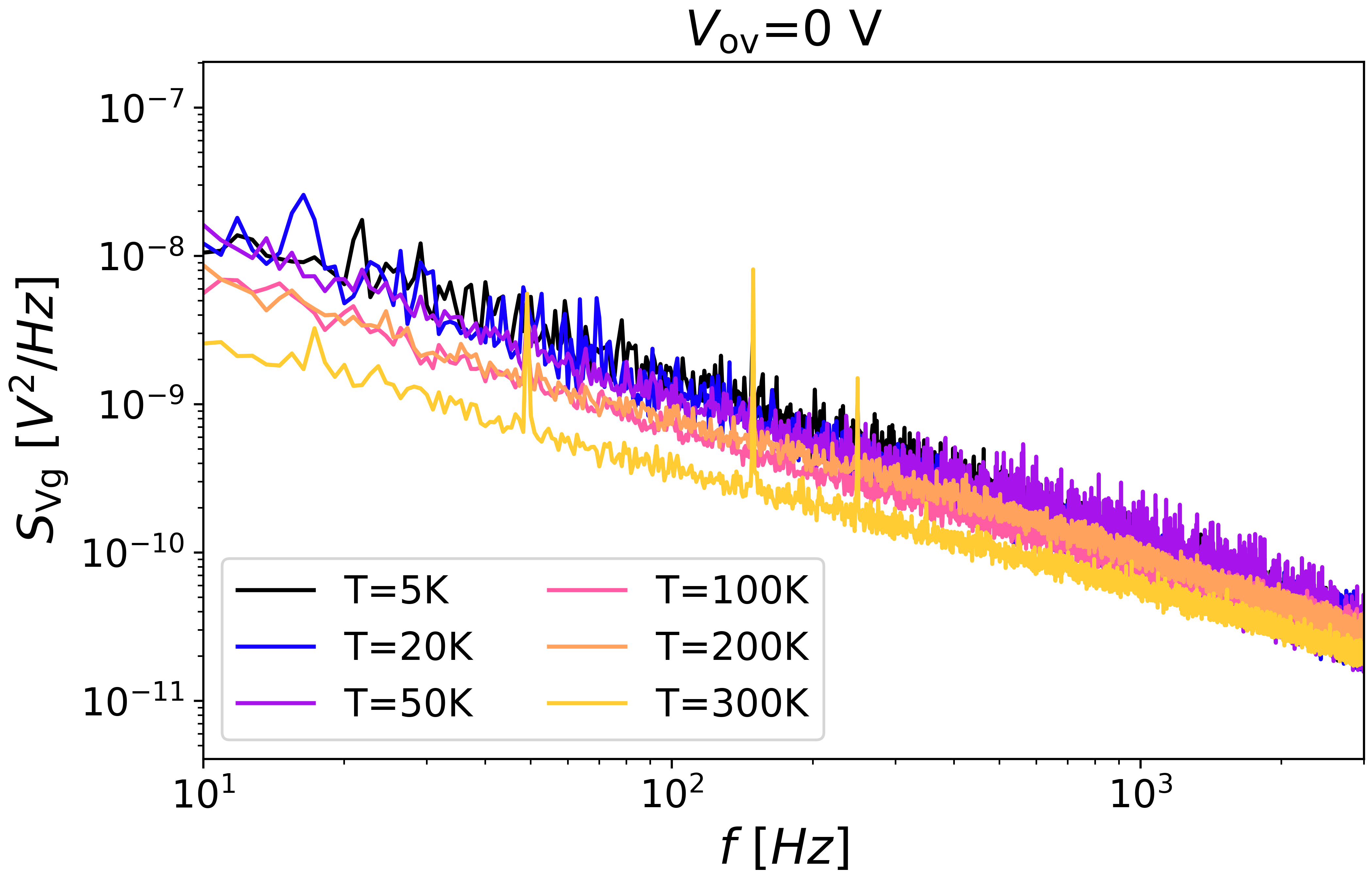}} \\
		\subfloat[][\label{fig:SVg_Vov}]
		{\includegraphics[width=.45\textwidth]{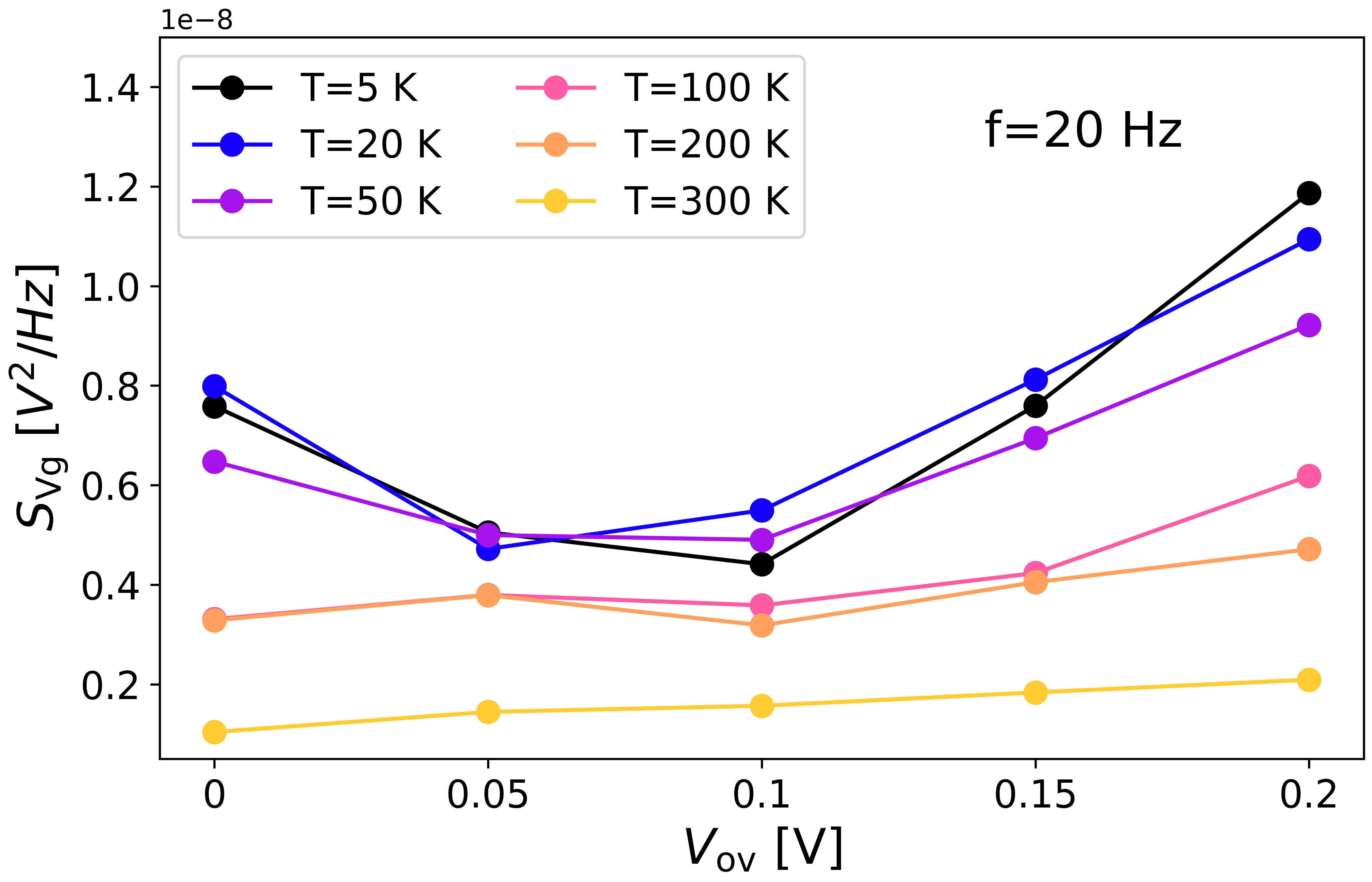}} \quad
		\subfloat[][\label{fig:SVg_branch}]
		{\includegraphics[width=.45\textwidth]{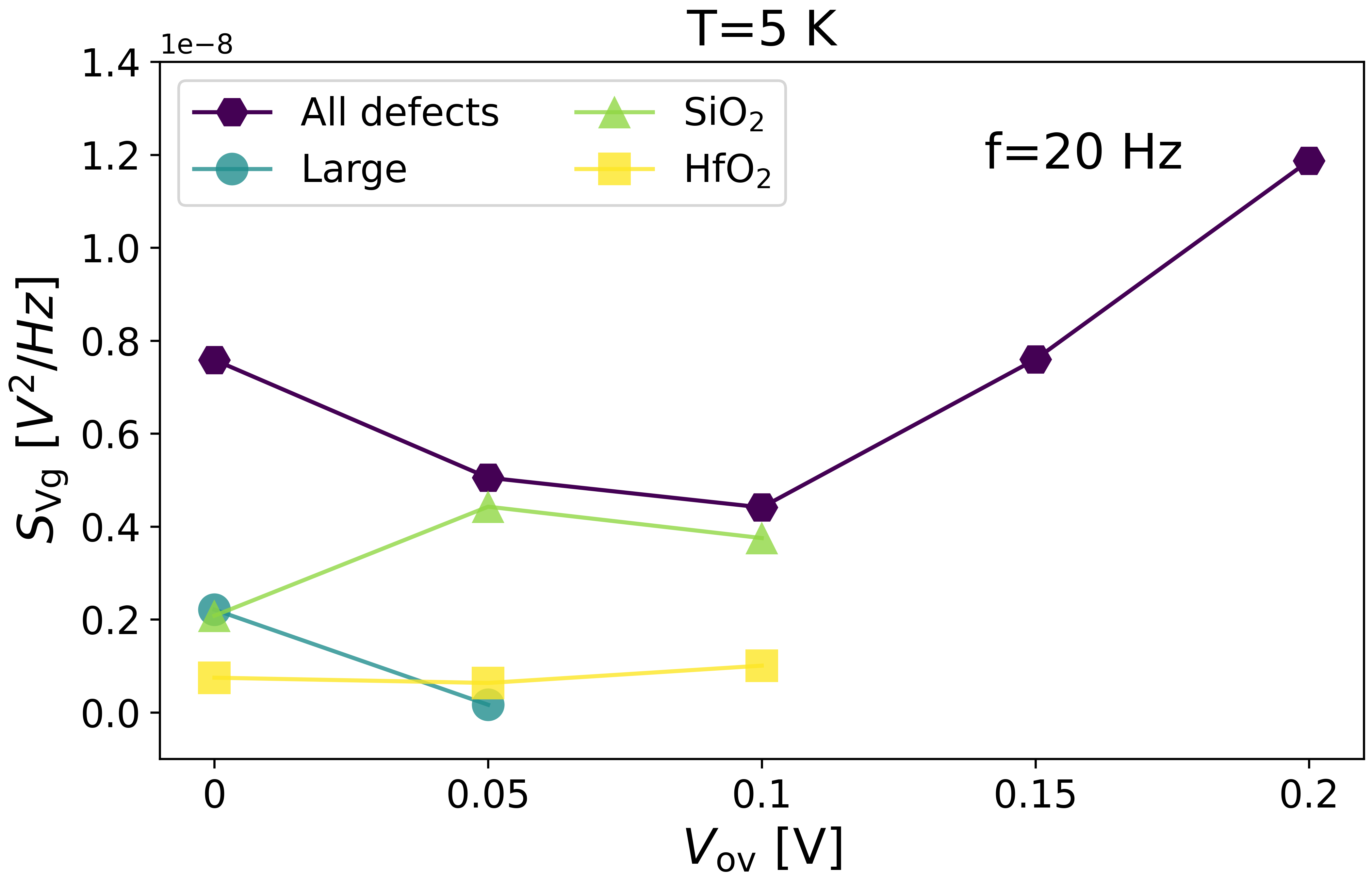}} \\
		\caption{\textbf{Fig.4} (a) Comparison between the noise power $N_\mathrm{P}$ calculated by considering all the measured traces and by considering just the traces with detectable RTN. $N_\mathrm{P}$ values are comparable, meaning that the background noise of the smart array is negligible, and most of the noise comes from the detected RTN traces. (b) $S_\mathrm{Vg}$ of the entire array across temperature. The spectra are the linear mean of all the power spectral densities of individual RTN traces. The resulting spectra are 1/f, even at T=5 K. The noise increase as the temperature is lowered, consistently with what has been measured on large area devices. (c) $S_\mathrm{Vg}(V_\mathrm{ov})$ for different temperatures. The temperature trend at $V_\mathrm{ov}=0 V$ is due to a larger percolation in this bias regime, since the channel is not full formed. With increasing $V_\mathrm{ov}$, percolation effects are mitigated, and $S_\mathrm{Vg}$ decreases. (d) Contribution of different defects branches to $S_\mathrm{Vg}$ at T=5 K. As $V_\mathrm{ov}$ increases, $\mathrm{SiO_2}$ defects become the main source of noise. Indeed, on one hand, higher overdrives reduce percolation effect, and on the other hand the defects bands probed change with $V_\mathrm{ov}$.}
		\label{fig:Fig4}
	\end{figure}
	
	1/f noise represents the spectral-domain manifestation of multiple RTN defects fluctuating simultaneously, with time constants following a log-normal distribution. Indeed, the power spectral density (PSD) of a single defect RTN has a Lorenztian shape, and summing the contribution of many defects whose time constants are log-normally distributed results in a 1/f spectrum \cite{grasser_noise_2020}. Based on this consideration, PSD of converted $\Delta V_\mathrm{th}$ single raw traces were calculated and summed over. Afterwards, the sum of all the PSDs has been divided by the total number of device measured. The LFN spectra obtained must be regarded as the $average$ noise of the entire device ensemble. In variability studies, the geometric (or logarithmic) mean is sometimes used to estimate the average noise, as it is less sensitive to the performance of devices in the tails of the statistical distribution, i.e., outliers \cite{grasser_noise_2020, ioannidis_impact_2014}. In the present work, however, the aim is to investigate the contribution to the 1/f noise of all detected defects, including those located in the tail of the trimodal exponential distribution shown in \cref{fig:CCDF 300K}. \\
	
	\noindent Before discussing the details of the LFN analysis, it is worth to point out that the noise coming from both the digital circuitry of the array and the experimental setup does not affect the reconstructed 1/f spectra substantially. To verify that, the noise calculated by taking into account just those traces which showed RTN was compared with the noise obtained using all the measured traces, which are almost three times more. In \cref{fig:Np} is shown the noise power $N_\mathrm{P}$ for sake of clarity. The noise level does not change significantly, meaning that the 1/f spectra are dominated by the detected devices' RTN. Furthermore, \cref{fig:Np} also provides a positive feedback on the RTN extraction method previously applied, showing that most of defects activity has been correctly detected.\\
	
	\noindent The input-referred 1/f drain current noise $S_\mathrm{Vg}$ follows a 1/f behaviour (see \cref{fig:SVg}), consistent with the commonly accepted theories, and it increases at cryogenic temperatures, as widely reported in literature \cite{asanovski_understanding_2023, cardoso_paz_performance_2020, oka_origin_2023, kiene_cryogenic_2024}. The value of $S_\mathrm{Vg}$ extracted at $f=20$ Hz is plotted in \cref{fig:SVg_Vov} as function of the overdrive. At $V_\mathrm{ov}=0$ V, $S_\mathrm{Vg}$ is higher at low temperatures because of both larger percolation and the activity of defects associated with the large branch, as discussed later. Increasing the gate voltage, the channel gets stronger, and $S_\mathrm{Vg}$ at cryogenic temperatures decreases, which is consistent with the vanishing of the third branch in the CCDF (see \cref{fig:CCDF 5K}). At $V_\mathrm{ov} > 0.1$ V, $S_\mathrm{Vg}$ slightly increases, especially at $\mathrm{T < 100}$ K. This trend could be attributed to a shift in the defect active energy area scanned by the channel Fermi level. Indeed, as $V_\mathrm{ov}$ increases, the Fermi level probes an increasing number of defects in the $\mathrm{SiO_2}$ layer, while traps in the $\mathrm{HfO_2}$ move outside the energy window swept by the Fermi level. This interpretation is supported by the modality of the CCDFs discussed in the previous section, which become monomodals at $V_\mathrm{ov} > 0.1$ V, and also by the trend in \cref{fig:SVg_branch}, where the contribution to $S_\mathrm{Vg}$ of defects associated to different modes is investigated. At $V_\mathrm{ov}=0$ V and $\mathrm{T=5}$ K, bulk $\mathrm{HfO_2}$ defects still play a significant role in the LFN. Indeed, an examination of \cref{fig:SVg_Vov} and \cref{fig:SVg_branch} reveals that the $S_\mathrm{Vg}$ component related to $\mathrm{HfO_2}$ layer is comparable to, or even higher than, the total $S_\mathrm{Vg}$ observed at $\mathrm{300}$ K. This finding is consistent with the analysis discussed in the previous section, where it was shown that more than 80\% of the total RTN activity came from the $\mathrm{HfO_2}$ layer. By inspection of \cref{fig:Prob} and \cref{fig:SVg_branch}, one can conclude that $\mathrm{HfO_2}$ defects are the main responsible for the high RTN activity at cryogenic temperatures and at $V_\mathrm{ov}=0$ V, but their contribution to LFN is small if compared to that of defects associated with the other branches, which contribute almost equally to the total $S_\mathrm{Vg}$. This opposite trend is consequence of the fact that the amplitude of the PSD of the original RTN traces is directly proportional to $\Delta V_\mathrm{th}$, which is larger for the interface traps. At $V_\mathrm{ov}=0.05$ V, the CCDF is still trimodal, but the contribution to the LFN of the third mode is mitigated. At $V_\mathrm{ov}=0.05$ V and $V_\mathrm{ov}=0.1$ V, most of the noise is attributed to $\mathrm{SiO_2}$ defects, whereas the contribution of the $\mathrm{HfO_2}$ branch is almost unchanged from $V_\mathrm{ov}=0$ V. At $V_\mathrm{ov} > 0.1$ V, CCDFs are monomodal and $S_\mathrm{Vg}$ is fully associated with traps in the $\mathrm{SiO_2}$ band. \\
	 
\begin{figure}
		\centering
		\captionsetup{labelformat=empty}
		\subfloat[][\label{fig:kT bell}]
		{\includegraphics[width=.45\textwidth]{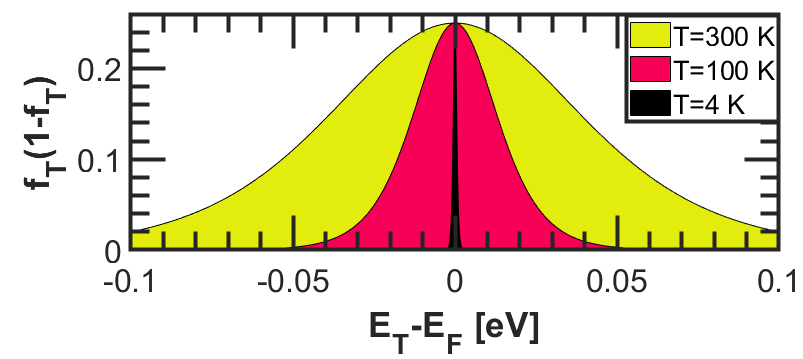}} \quad
		\subfloat[][\label{fig:n_tot}]
		{\includegraphics[width=.45\textwidth]{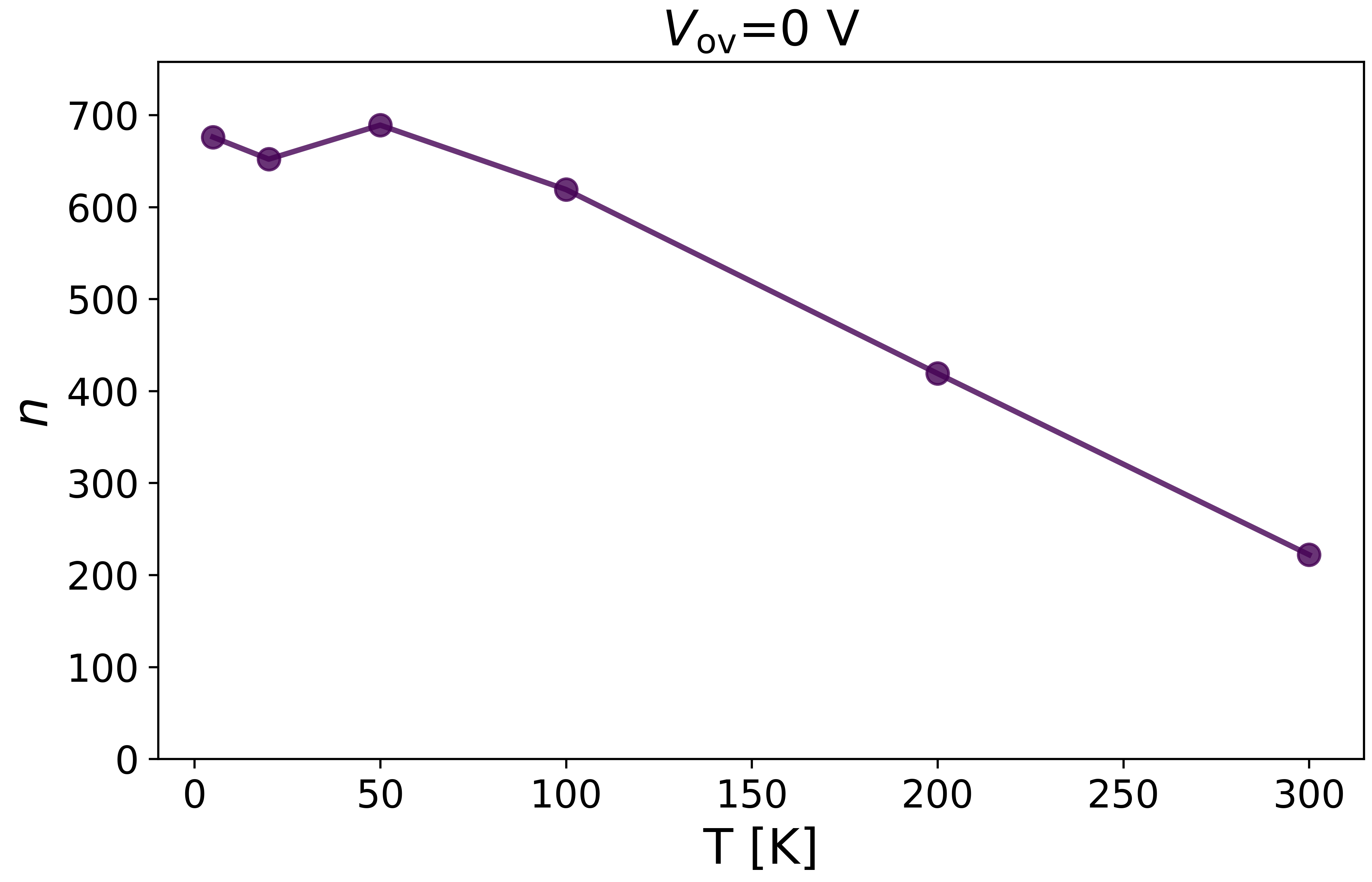}} \\
		\caption{\textbf{Fig.5} (a) Plot of $f_\mathrm{T}(1-f_\mathrm{T})$ versus $E_\mathrm{T}-E_\mathrm{F}$ for different temperatures. The coloured area under the curves represents the integral and it is equal to $k_B\mathrm{T}$. The figure is taken from \cite{asanovski_understanding_2023}. (b) The total number of detected defects $n$ increases rapidly at $\mathrm{T} < 300$ K. }
		\label{fig:Fig5}
	\end{figure}

	\noindent It is worth to further investigate some consequences of the previous discussions on the possible oxide defects origin. Since the gate voltages considered here are around $V_\mathrm{th}$, LFN is assumed to be in the carrier number fluctuation (CNF) regime. $S_\mathrm{Vg}$ can then be written as
	
	\begin{equation}
		\label{eqn:SVg}
		S_\mathrm{Vg}=\frac{qk_\mathrm{B}TN_\mathrm{BT}}{WLC_\mathrm{ox}^2 \alpha}\cdot\frac{1}{f}
	\end{equation} 
	\\
	\noindent where $k_\mathrm{B}$ is the Boltzmann constant, $N_\mathrm{BT}$ is the volumetric gate dielectric trap density per unit energy, $C_\mathrm{ox}$ is the oxide capacitance per unit area, $f$ is the frequency and $\alpha$ is the wentzel-kramers-brillouin (WKB) tunnelling probability for a rectangular barrier. The $k_\mathrm{B}\mathrm{T}$ factor in \text{eq.}\eqref{eqn:SVg} originates from the integral $\int_{-\infty}^{\infty} f_\mathrm{T}(1-f_\mathrm{T}) dE_\mathrm{T}$, being $E_\mathrm{T}$ the trap energy and $f_\mathrm{T}$ the near-equilibrium occupation function of the defect, which is described by a Fermi function with Fermi level $E_\mathrm{F}$ \cite{asanovski_understanding_2023}. The integral above is then the area of the bell-shaped function centred at $E_\mathrm{T}=E_\mathrm{F}$ (see \cref{fig:kT bell}). In other words, only defects with energies within a few $k_B\mathrm{T}$ of the Fermi level can contribute to the noise. This implies that, as shown in \cref{fig:kT bell}, at $\mathrm{T=4.2}$ K the active energy area is almost two orders of magnitude smaller than at $\mathrm{T=300}$ K. On the other hand, the total number of defects showing RTN and, therefore, contributing to LFN, increases dramatically at cryogenic temperatures (see \cref{fig:n_tot}). One possible way to reconcile these results is to postulate that the defect density increases at low temperatures. As already stressed, at $V_\mathrm{ov} \leqslant \mathrm{0.05}$ V, most of active defects are in the $\mathrm{HfO_2}$ layer. Hence, the following discussion focuses on the physical origin of electron trapping in this material. Several works \cite{izmailov_electron_2021, izmailov_shallow_2021, izmailov_electron_2024, strand_intrinsic_2018} showed that electron trapping in $\mathrm{HfO_2}$ is related to disorder, which creates electron traps associated with low-coordinated sites and elongated $\mathrm{Hf-O}$ bonds. The trapping is followed by significant polaron-like distortion of the surrounding amorphous network. At cryogenic temperatures, due to the limited thermal energy available, an electron has a significantly higher probability of remaining near the precursor site in the near-interface oxide long enough to polarize the oxide network, resulting in a shallow trapping event. Afterwards, the electron can either tunnel into this self-created trap state or remain nearby without contributing to the channel current. \\ 
	
	\noindent As already discussed, at $V_\mathrm{ov} > 0.1$ V, $S_\mathrm{Vg}$ is primarily associated with $\mathrm{SiO_2}$ defects due to shift in the active energy area swept by the Fermi level. The physical origin of the kinetics of these defects, which do not freeze out, is still under debate. Recently, in was shown that a full quantum-mechanical solution of the non-radiative multi-phonon (NMP) model can interpret the experimental results at $\mathrm{T=4.2}$ K \cite{michl_efficient_2021, michl_evidence_2021, michl_efficient_2021-1}. According to this solution, nuclear tunnelling becomes the predominant mechanism of charge trapping at cryogenic temperatures. Beside that, the number of detected defects in the $\mathrm{SiO_2}$ layer at higher overdrives is weakly temperature dependent (see Supplement Material), which is challenging to reconcile with the previous discussion based on the trend shown if \cref{fig:kT bell}. Further analyses are required to comprehend the nature of charge trapping in $\mathrm{SiO_2}$ at cryogenic temperatures. 
	
	

	\subsection{Time constants and step heights correlation}\label{sec5}

	\begin{figure}
		\centering
		\captionsetup{labelformat=empty}
		\subfloat[][\label{fig:correlations_300K}]
		{\includegraphics[width=.45\textwidth]{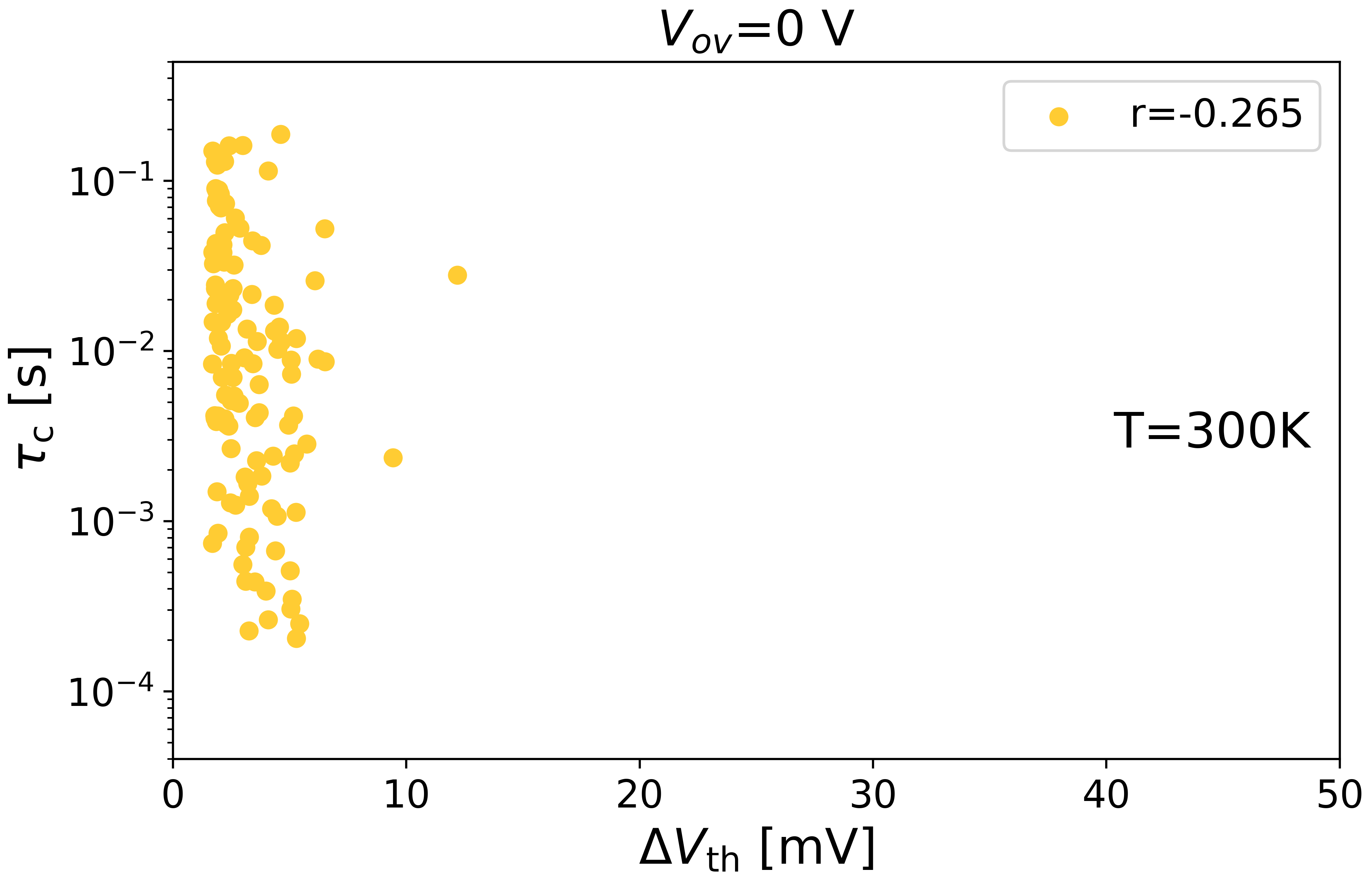}} \quad
		\subfloat[][\label{fig:correlations_emissions_300K}]
		{\includegraphics[width=.45\textwidth]{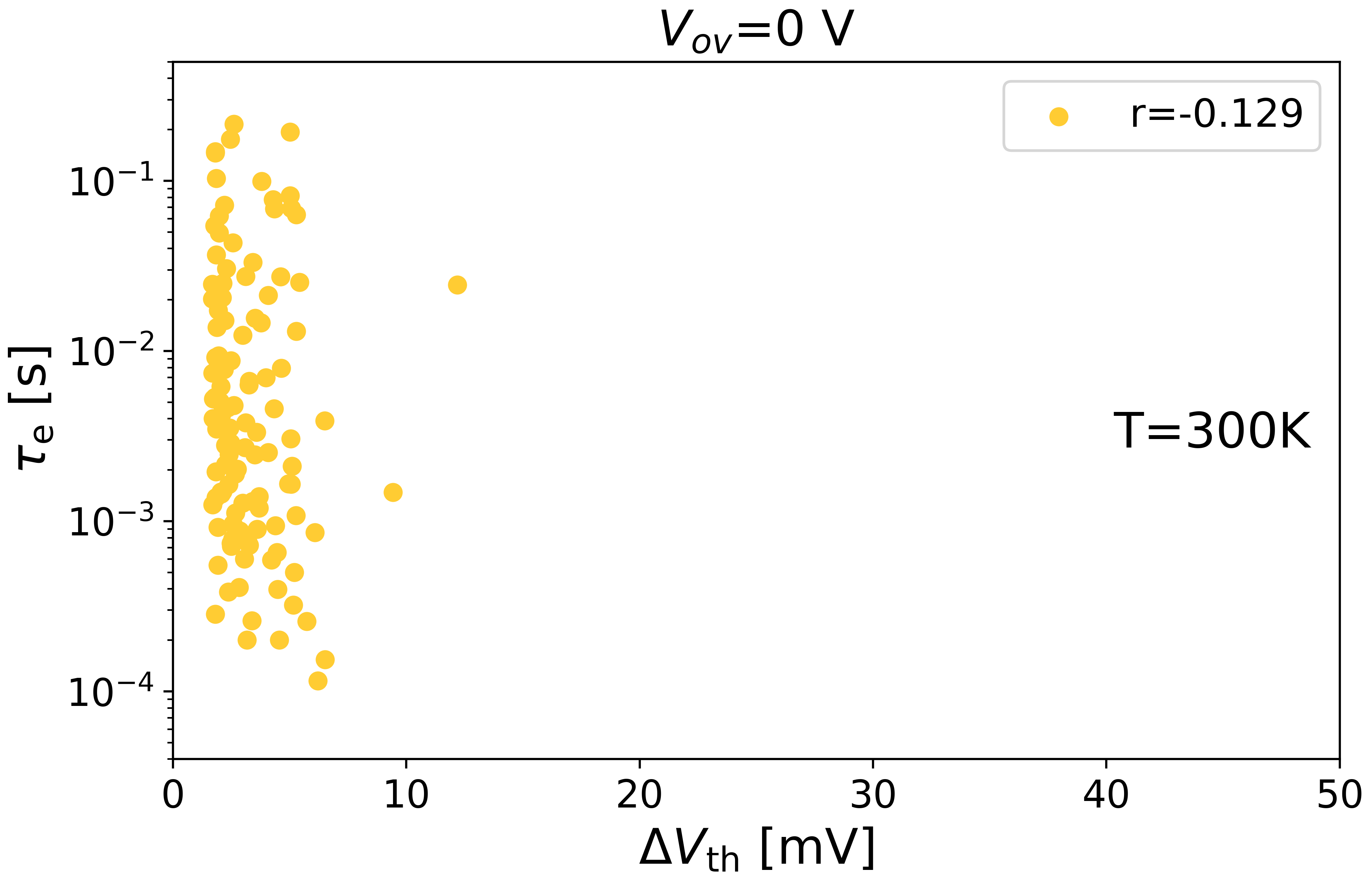}} \\
		\subfloat[][\label{fig:correlations_5K}]
		{\includegraphics[width=.45\textwidth]{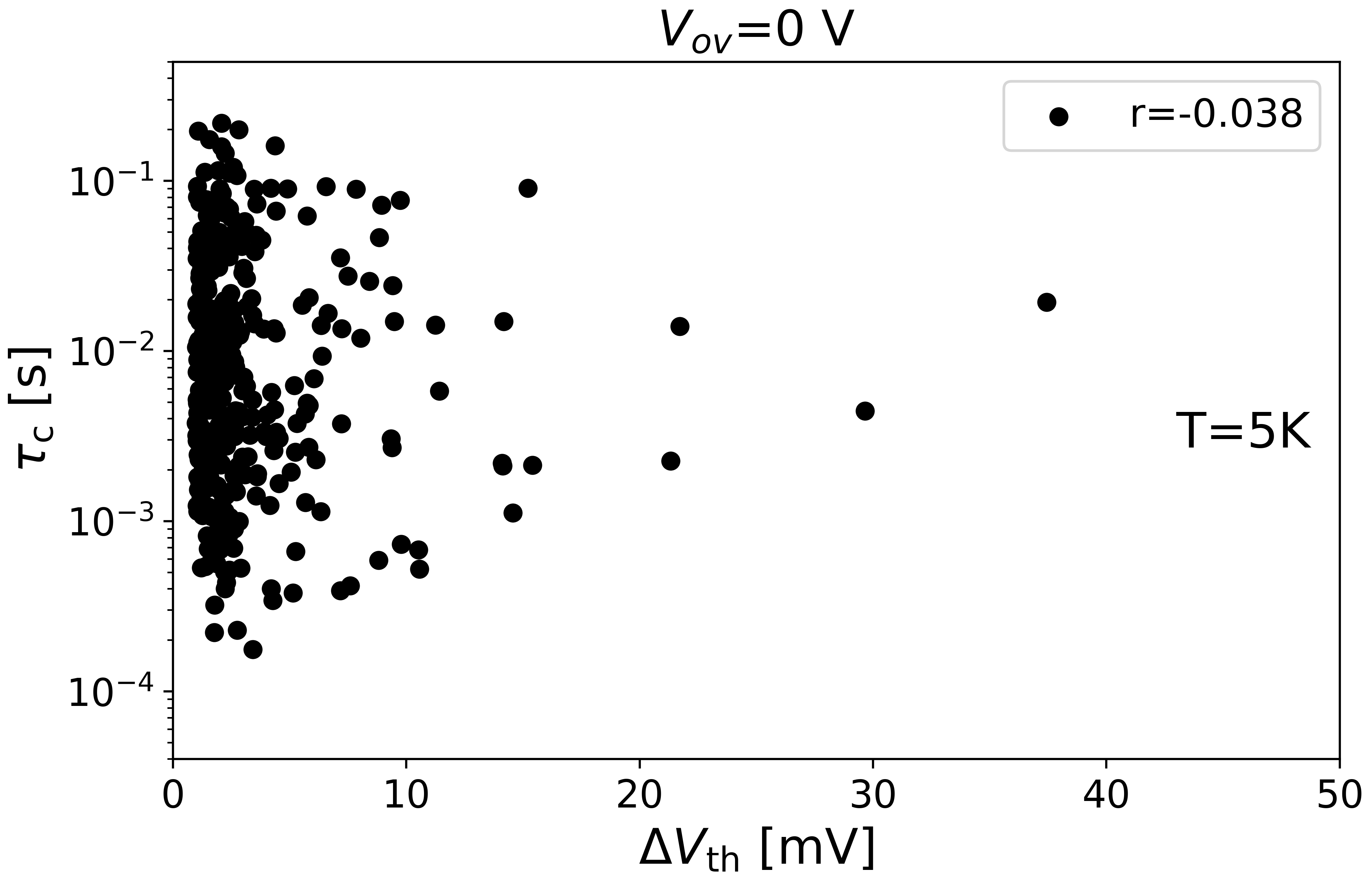}} \quad
		\subfloat[][\label{fig:correlations_emissions_5K}]
		{\includegraphics[width=.45\textwidth]{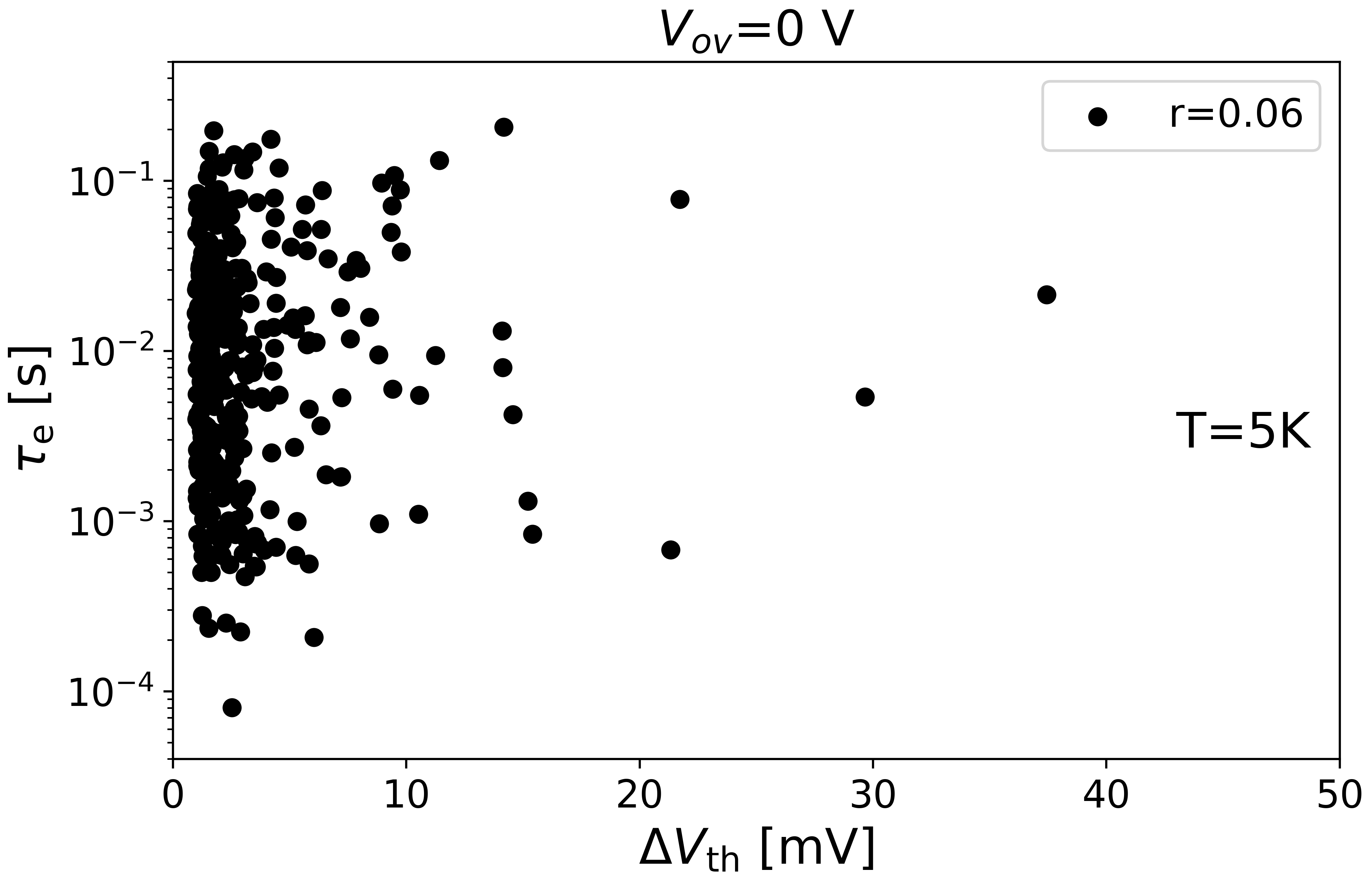}} \\
		\caption{\textbf{Fig.6} Scatter plot of trap step height $\Delta V_\mathrm{th}$ and capture time constant $\tau_\mathrm{c}$ at T=300 K (a) and T=5 K (b) and $V_\mathrm{ov}=0$ V. At both temperatures, no correlation is observed. Similar result is obtained by plotting $\Delta V_\mathrm{th}$ against emission time constant $\tau_\mathrm{e}$ (c)-(d). These results prove that McWorther model is not accurate to describe charge trapping in MOSFET oxides. Indeed, it assumes elastic tunnelling between the channel and the defect, which would entail a correlation between the defect position, and thus its step height, and its time constants.}
		\label{fig:Fig6}
	\end{figure}

	McWhorter model, which is based on elastic tunnelling, predicts an exponential relationship between defect capture (emission) time constant $\tau_\mathrm{c,e}$ and its position in the oxide $x_\mathrm{T}$ \cite{grasser_stochastic_2012}:
	
	\begin{equation}
		\label{eqn:tau}
		\tau_\mathrm{c,e} \propto e^{\frac{x_\mathrm{T}}{t_\mathrm{ox}}}
	\end{equation} 
	\\ 
	
	\noindent Since $x_\mathrm{T}$ determines the defect amplitude, if charge trapping processes are dominated by elastic tunnelling, there must be a correlation between $\tau_\mathrm{c,e}$ and $\Delta V_\mathrm{th}$. Many works already proved the inadequacy of elastic tunnelling and, therefore, of McWhorter model to describe LFN in MOSFET devices at $\mathrm{T=300}$ K \cite{nagumo_statistical_2010, abe_understanding_2011, saraza-canflanca_statistical_2021}. Here, the possible correlation between $\tau_\mathrm{c,e}$ and $\Delta V_\mathrm{th}$ is investigated from $\mathrm{T=300}$ K down to $\mathrm{T=5}$ K.  It is worth to stress that, since the $\Delta V_\mathrm{th}$ depends $linearly$ with the trap position and, according to \text{eq.}\eqref{eqn:tau}, $\tau_\mathrm{c,e}$ depends $exponentially$ with the trap position, a linear correlation on a semi-logarithmic axis is expected. In \ref{fig:correlations_300K} and \ref{fig:correlations_emissions_300K} the scatter plots $\Delta V_\mathrm{th} - \tau_\mathrm{c,e}$ at $\mathrm{T=300}$ K are reported, together with the calculated Pearson coefficient $\rho$. No correlation is found, in agreement with \cite{nagumo_statistical_2010, abe_understanding_2011, saraza-canflanca_statistical_2021}. Decreasing temperature, results are similar (see Supplement Materials), down to $\mathrm{T=5}$ K (\ref{fig:correlations_5K} and \ref{fig:correlations_emissions_5K}). From these results it is evident that McWhorter model is not accurate in describing LFN and that defect kinetic is not dominated by the elastic tunnelling, neither at cryogenic temperatures. 
	
	\section{Conclusion}\label{sec6}
	
	In this work, statistical random telegraph noise characterization has been employed to explore the physical origin of 1/f noise in cryoCMOS n-type devices. \\
	\noindent The first part of the paper was devoted to the analysis of the threshold voltage shifts due to charged defects, reported in cumulative plots. In this context, it was shown that, at $\mathrm{T = 300}$ K, the CCDF followed a monomodal exponential distribution, and only defects associated with $\mathrm{SiO_2}$ layer contributed to RTN. At cryogenic temperatures, instead, the CCDF changed its modality with the appearing of other two modes: a small one, attributed to $\mathrm{HfO_2}$ defects, and a large one whose origin is still controversial. An interpretation of both the trimodal shape of the CCDF and the nature of the third branch was provided exploiting the percolation theory. The change in the shape of the CCDF at cryogenic temperatures was attributed to a major sensitivity of the inversion carriers to channel inhomogeneities. Indeed, even though random dopants and surface potential fluctuations are the main cause of the exponential distribution of the $\Delta V_\mathrm{th}$ already at $\mathrm{T=300}$ K, at low temperature carriers' thermal energy is reduced and they are more susceptible to channel disorder. This larger non-uniformity can induce the formation of insulating regions in the channel, increasing the electrostatic impact of oxide and interface defects: this could origin the third mode of large $\Delta V_\mathrm{th}$ experimentally observed. Nonetheless, the primary outcome of the first part of this study is the identification of electron trapping within the $\mathrm{HfO_2}$ layer as the dominant contributor for over 80\% of the total RTN measured at $\mathrm{T = 5}$ K. \\
	\noindent The second half of this work is dedicated to low frequency noise. The aim is to investigate the contribution of individual defect branches to the total 1/f noise, reconstructed using the original time traces. Even though most of active defects at cryogenic temperatures are associated with the $\mathrm{HfO_2}$ layer, their electrostatic impact is quite small if compared to interface traps. Therefore, most of low frequency noise arises from interfaces defects. Few defects cause most of the 1/f noise at cryogenic temperatures. Nevertheless, LFN generated by bulk $\mathrm{HfO_2}$ traps at $\mathrm{T = 5}$ K is still comparable to the total LFN at $\mathrm{T = 300}$ K. An interpretation of this huge activity of bulk defects was proposed and it was based on the concept of electron self-localization due to the polarization of the $\mathrm{HfO_2}$ network. With increasing overdrive voltage, the noise contribution from $\mathrm{HfO_2}$ traps steadily diminishes until it becomes negligible.\\
	\noindent In the final part of the paper the correlation between defect time constants and threshold voltage shifts was investigated. This correlation is expected according to the elastic tunnelling model and, consequently, the McWorther theory. However, no correlation was found, either at room or at cryogenic temperatures.

	\section{Methods}
	\subsection{Smart Array}
	All the tests presented in this work were conducted using our transistor-array chip, fabricated in a commercial 28 nm HKMG technology.\\
	
	\noindent To efficiently utilize the chip area in our array, the design includes a total of 2500 addressable CORE transistors distributed across 10 rows and 250 columns. Each column contains 10 transistors connected by their gate terminals, while each drain is connected to an individual drain line (see Figure 1a). In this design, groups of 10 transistors, i.e., each column, can be selected for electrical operation via a row and column selection circuitry. This circuitry consists of two-layer shift registers embedded in an individual decoder present in each selectable column. The selection circuitry enables individual or multiple column activations for electrical measurements through a 260-bit selection word, comprising a 10-bit row selection and a 250-bit column selection.\\
	
	\noindent Voltage application to the selected column is performed via a decoder circuit that processes the row and column bits for each column. It allows for the establishment of two different voltages, namely $V_\mathrm{g,ON}$ or $V_\mathrm{g,OFF}$, via separate full pass gates constructed with thick IO devices to prevent unwanted degradation during above-nominal voltage application. Device electrical characterization is performed using $V_\mathrm{g,ON}$, which enables the execution of time-dependent variability effects such as RTN, BTI, or HCD. This voltage can range from 0 V up to 2.5 V. Meanwhile, $V_\mathrm{g,OFF}$ is used to bias the non-selected columns, typically at a negative voltage, to reduce off-state leakage.\\
	
	\noindent The chip design also incorporates sense lines at the output of the shift register, as well as on the gate and drain of a device within the array. This allows us to verify the functionality of the periphery and identify potential voltage drops on the gate and drains. Moreover, all measurements are performed using a custom-built switching matrix with separately controllable voltage sources to manage the periphery, along with K2600 SMUs for current readouts.
	
	\subsection{Extended Log Likelihood formulation for unknown amount of left censored steps}
	\noindent The regularized formulation of the Likelihood function involves a Multinomial prefactor in terms of the number of steps hypothesized.
	In many maximum likelihood fitting contexts, the sample size is a constant in terms of the distribution parameters fitted, so that it can be dropped from the maximization. When an unknown amount of left censored steps needs to be included in the fit, however, this prefactor is not a constant anymore. The more generic expression for the Log Likelihood function in terms of the (multimodal) step distribution with PDF $f[\Delta V_\mathrm{th},\vec{n},\vec{\eta}]$ needs to be considered then:
	
	\begin{multline}
		\label{eqn:multimodal}
		 \mathrm{PDF}[\mathrm{MultiModalDistribution}[n_\mathrm{Tot}, \{F[\Delta V_\mathrm{th,d}, \vec{n},\vec{\eta}]^{n_\mathrm{Tot}-n_\mathrm{d}}, \\ f[\Delta   V_\mathrm{th,1}, \vec{n},\vec{\eta}],...,f[\Delta V_\mathrm{th,n_d}, \vec{n},\vec{\eta}]\}], \{n_\mathrm{Tot}-n_\mathrm{d}, 1..., 1\}] = \\
		  \frac{n_\mathrm{Tot}!}{(n_\mathrm{Tot}-n_\mathrm{d})!} F[\Delta V_\mathrm{th,d}, \vec{n},\vec{\eta}]^{n_\mathrm{Tot}-n_\mathrm{d}} \prod_{k=1}^{n_\mathrm{d}} 
		  f[\Delta V_\mathrm{th,k},\vec{n},\vec{\eta}]
	\end{multline} 
	\\ 
	
	\noindent with PDF
	
	\begin{equation}
		f[\Delta V_\mathrm{th},\vec{n},\vec{\eta}] = \frac{\vec{n}}{n_\mathrm{Tot}} \cdot \Big(\frac{e^{-\frac{\Delta V_\mathrm{th}}{\vec{\eta}}}}{\vec{\eta}}\Big)
	\end{equation}
	\\
	
	\noindent The binomial prefactor in terms of the unknown total amount of steps $n_\mathrm{Tot}$ has to be absorbed in the Log Likelihood expression to be maximized:
	
	\begin{multline}
		\Lambda[\vec{\Delta V_\mathrm{th}},\vec{n},\vec{\eta}] = \\ \mathrm{Log(\Gamma}[n_\mathrm{Tot} + 1]) - \mathrm{Log(\Gamma}[n_\mathrm{Tot} - n_\mathrm{d} + 1])
		+ (n_\mathrm{Tot} - n_\mathrm{d})\mathrm{Log}[F[\Delta V_\mathrm{th,d}, \vec{n}, \vec{\eta}]] + \sum_{k=1}^{n_\mathrm{d}}\mathrm{Log}[f[\Delta V_\mathrm{th,k},\vec{n},\vec{\eta}]]
	\end{multline}
	
	\subsection{Suppress over-fitting through constrained maximum likelihood fitting over 6 temperatures simultaneously}
	\noindent By performing a constrained ML fit over the 6 temperatures simultaneously, the impact of over-fitting can be suppressed: imposing a monotonously descending – which is an extra assumption of course – $\eta[\mathrm{T}]$ dependence for each mode where it is observable drives the fit towards more realistic values for the distribution parameters.
	
	\subsection{Identifying presence of extra step size distribution modes using outlier detection hypothesis test}
	\noindent Starting from monomodal Exponential distribution fits, the presence of extra observable step size distribution modes for each of the CCDF curves fitted can be assessed using an outlier testing hypothesis test, either for the independent unconstrained ML fits, either for the constrained ML fits onto the combined CCDF curves at each of the 6 temperatures measured. The probability of finding the larger value $x_n$ in the sample observed given the distribution modality and distribution parameters estimated can be applied as a significance test: 
	
	\begin{equation}
		P[x_1 < x_n]\mathrm{\&}P[x_2 < x_n]\mathrm{\&}...\mathrm{\&}P[x_n < x_n] = F[x_n]^n
	\end{equation}
	
	\noindent To avoid subtractive cancellation, it's safer to compute the tail of the distribution fitted in terms of the test significance probability, and to use the matrix exponential function to maintain precision:
	
	\begin{equation}
		exp[nLnp1[-R[x_n]]]<1-F_\mathrm{Signif}
	\end{equation}
	
	\noindent As soon as the tail probability exceeds the critical value for the significance probability $F_\mathrm{Signif}$ chosen, $x_\mathrm{n}$ should be considered an outlier for the distribution hitherto fitted. When adhering to the hypothesis of underlying exponential distributions, the presence of an extra exponential mode can then be inferred as an alternate hypothesis, whereupon the ML fit can be repeated accordingly. This process can then be repeated after a bimodal fit to detect the presence of a third (i.c. typically larger) step mode.
	
	\bibliography{Bib.bib} 

\end{document}